\documentclass[12pt,preprint2]{emulateapj}

\usepackage{pslatex}
\usepackage[T1]{fontenc}
\usepackage[latin1]{inputenc}
\usepackage{subfigure}
\usepackage{amsmath}
\usepackage{natbib}
\usepackage{graphicx,graphics}
\usepackage{amssymb}
\usepackage[english]{babel}

\def\mh{M_{\bullet}}
\def\mbh{m_{\rm BH}}

\def\rh{r_{\rm infl}}

\def\rc{r_{\rm core}}
\newcommand{\gap}{\;\rlap{\lower 2.5pt \hbox{$\sim$}}\raise 1.5pt\hbox{$>$}\;}
\newcommand{\lap}{\;\rlap{\lower 2.5pt \hbox{$\sim$}}\raise 1.5pt\hbox{$<$}\;}
\newcommand{\beq}{\begin{equation}}
\newcommand{\eeq}{\end{equation}}
\newcommand{\msun}{M_\odot}

\newcommand{\sgr}{Sgr${\rm A}^*$}

\shorttitle{Stars and Remnants at the Galactic Center}
\shortauthors{David Merritt}

\begin{document}

\title{The Distribution of Stars and Stellar Remnants at the Galactic Center} 

\author{David Merritt}
\affil{Department of Physics and Center for Computational Relativity and 
Gravitation, Rochester Institute of Technology, Rochester, NY 14623, USA}

\begin{abstract}
Motivated by recent observations that suggest a low density of old
stars around the Milky Way supermassive black hole, 
models for the nuclear star cluster are considered that have
not yet reached a steady state under the influence of gravitational
encounters.
A core of initial radius $1-1.5$ pc evolves to a size of approximately
$0.5$ pc after $10$ Gyr, roughly the size of the observed core.
The absence of a Bahcall-Wolf cusp is naturally explained in these
models, without the need for fine-tuning or implausible initial conditions.
In the absence of a cusp, the time for a $10\msun$ black hole to 
spiral in to the  Galactic center from an initial distance of $5$ pc
can be much greater than $10$ Gyr.
Assuming that the stellar black holes had the same phase-space distribution
initially as the stars, their density after $5-10$ Gyr is predicted
to rise very steeply going into the stellar core, but could remain
substantially below the densities inferred from steady-state
models that include a steep density cusp in the stars.
Possible mechanisms for the creation of the parsec-scale initial core
include destruction of stars on centrophilic orbits 
in a pre-existing triaxial nucleus, inhibited star formation near
the supermassive black hole, or ejection of stars by a massive
binary.
The implications of these models are discussed for the rates of 
gravitational-wave inspiral events, as well as other physical 
processes that depend
on a high density of stars or stellar mass black holes near \sgr.
\end{abstract}


\section{Introduction}

Near-infrared imaging reveals a 
cluster of stars around the $\sim 4\times 10^6\msun$ 
supermassive black hole (SMBH) at the center of the Galaxy.
Much recent work has focussed on the young stars
that dominate the total light in the central parsec
\citep{Forrest:87,Allen:90,Krabbe:91}.
These stars have masses of $10-60\msun$ and appear to have formed
in one or more starbursts during the last few million years 
\citep{Krabbe:95,Paumard:06}.
While their total numbers are small, the density of the young
stars increases very steeply toward the SMBH \citep{Genzel:03,Paumard:06}.
How such massive stars can form so deep in the SMBH potential well
remains an unsolved problem \citep{Alexander:05,Paumard:08}.

The dominant population at the Galactic center consists
of old stars: mostly metal-rich, M, K and G type giants with
masses of one to a few Solar masses \citep{Blum:03,Davies:09}.
The late type giants dominate the total flux outside the central
parsec, 
and seeing-limited observations as early as the 1960s showed that
their surface density follows a power law, 
$\Sigma\sim R^{-1}$ at $R\lap 10$ pc, 
implying a space density profile $n\sim r^{-2}$
\citep{BN:68,McGinn:89,Haller:96}.
Inside $\sim 0.5$ pc, a drop in the CO absorption strength 
\citep{Sellgren:90,Haller:96}
signals a decrease in the projected density of old stars
\citep{Genzel:96,Figer:00,Scoville:03,Figer:03}.
However the possibility of contamination by light from early
type stars at these radii made such inferences uncertain.

Three recent studies have clarified the situation.
\citet{BSE:09} took deep narrow band images of the inner
parsec and used a CO absorption feature to distinguish early
from late type stars.
They classified roughly 3000 stars down to a $K$-band magnitude
of $\sim 15.5$.
Of these roughly 300 were early types, distributed as a steep
power-law  around the SMBH.
The late type stars showed a very different distribution:
a core of radius $\sim 0.5$ pc, with a constant or 
even declining surface density inside.
Do et al. (2009) obtained higher dispersion spectra of a smaller
sample of stars in a set of fields within 0.16 pc from
\sgr, estimated to be 40\% complete down to $m_K=15.5$.
They also found a flat or centrally-declining projected density 
of old stars.
\citet{Bartko:09b} found a flat distribution of late-type
stars with $m_k\le 15.5$ based on spectroscopic identifications
in a set of fields in the central $25''$.

Inferring the {\it space} density profile of the old stars
from the number counts in the inner parsec is difficult, since 
the projected density is apparently dominated by stars that are 
far from the center.
But the behavior of the number counts is difficult to reconcile
with models in which the space density increases
inwards.
Buchholz et al. (2009) found a best-fit power law of 
$\Sigma\propto R^{0.17}$ (projected density) for late type
stars inside $6''$ and $\Sigma\propto R^{-0.7}$ outside $6''$.
This corresponds to a space density that drops toward the center,
although the counting statistics were also consistent with
a flatter profile.
Do et al. (2009) inferred a steeper rate of central decline,
$\Sigma\propto R^{0.27}$, from their smaller data set, though
with larger uncertainties.
Do et al. compared the numbers counts with projections of power law
models in $n(r)$ and concluded 
that the density could not increase
faster than $n\sim r^{-1}$ toward the center,
although the preferred dependence was shallower.

As the authors of these three studies note, previous descriptions
of the stellar density as a broken power law,
with $n\sim r^{-1.2}-r^{-1.4}$
inside $\sim 0.4$ pc and $n\sim r^{-2}$ outside
\citep{Genzel:03,Schoedel:07},
were somewhat misleading, since the counts at small radii
were dominated by early type stars while the counts at large
radii were dominated by late type stars.
Presumably, the physical mechanisms responsible for creating
these two populations, and placing them on their current orbits,
were different and occurred at different times.

Models with a central ``hole'' naturally resolve one long-standing puzzle: why
virial estimates of the SMBH mass based on velocities
of late type stars 
\citep{Ghez:98,Genzel:00,CS:01,Eckart:02}
gave systematically lower values than the mass inferred
from the inner S-star orbits \citep{Gillessen:09,Ghez:08}.
If most of the old stars within the projected central parsec are
actually far from the SMBH, their motions will be 
relatively unaffected by its gravitational force,
yielding spuriously low virial masses.
Recent studies that explicitly incorporate a low central 
density for the late type stars \citep{Zhu:08,Schoedel:09}
yield virial masses for the SMBH that for the first time are
consistent with the mass obtained from the orbital fits.

However the low density of old stars at the 
Galactic center creates new puzzles.
A standard assumption has long been that stars near the SMBH should 
exhibit a cusp in the density, $n\sim r^{-\gamma}$, $\gamma\approx 7/4$.
This is the Bahcall-Wolf (1976) solution, which
holds for a relaxed (in the sense of gravitational encounters) 
population of stars moving in a point-mass potential.
A number of theoretical studies have argued that the Bahcall-Wolf solution is a robust
outcome, depending only weakly on the initial conditions and the
range of mass groups present \citep{MCD:91,Freitag:06a,HA:06a}.

It is possible that a Bahcall-Wolf cusp is present but that 
the stellar luminosity function changes inside $\sim 0.5$ pc,
such that the brightest stars are missing \citep{BSE:09,Do:09,Bartko:09b}.
This could  be a result of (physical)  collisions before or
during the red giant phase, which strip stellar envelopes and keep stars
from reaching their peak luminosities
\citep{Genzel:96,Alexander:99,BD:99};
or tidal interactions between single stars and the SMBH 
\citep{DK:05};
or it could result from an initial mass function that is truncated 
below $\sim 3\msun$, since these are the stars that would otherwise 
dominate the $K$-band number counts now \citep[e.g.][]{NS:05}.

While a ``hidden'' density cusp is possible,
it seems reasonable also to consider models in which
the observed stars are representative of the unobserved stars.
In these models, both populations would have a relatively low density in the
inner parsec.
Aside from Occam's principle,
a number of other motivations exist for considering such models:

1. {\it The relaxation time at the Galactic center is long.}
The stellar mass density at a distance of $2-3$ pc from \sgr 
-- roughly the SMBH  influence radius $\rh$ -- implies a two-body relaxation 
time (for Solar-mass stars) of 20-30 Gyr.
This is the relevant time scale for refilling of an evacuated
core if its initial radius is $\sim\rh$.

2. {\it Stellar kinematics suggest a low mass density in the inner parsec.}
Sch\"odel et al. (2009) found that proper motion data were consistent
with a range of models for the distribution of mass in the inner
parsec, but the best-fitting models had a flat or declining mass
density toward \sgr, similar to what is seen in the number counts of the 
old giants.

3. {\it Physical collisions fail, by a wide margin, to
predict the observed depletion of giant stars} in the faintest
magnitude bins.
Dale et al. (2009) concluded that giants in the $15>m_K>12$ magnitude
range would only be signficantly depleted within $\sim 0.01$ pc, even 
assuming a density of unseen colliders (main sequence stars, stellar-mass BHs) 
that was four times larger than in the dynamically relaxed models
with a Bahcall-Wolf cusp \cite[e.g.][]{HA:06a}.

4. Whatever their origin, {\it cores are ubiquitous components of 
galaxies with SMBHs,} at least in galaxies that are bright enough
or near enough for parsec-scale features to be resolved \citep{ACSIV}.
Core radii are roughly equal to SMBH influence radii
\cite[e.g.][]{Graham:04},
which in the case of the Milky Way would predict a core of radius 
$2-3$ pc.

This paper examines the viability of models for the Galactic center
that include a low-density core.
Core models are examined first from a structural point of view (\S 2),
then from the point of view of self-consistent equilibria (\S 3),
and finally from an evolutionary standpoint (\S 4).
The basic result is that core models ``work'': they reproduce
the number counts and kinematics of the late type stars, without
invoking physical collisions (collisions would be extremely
rare in these models) and without the necessity of fine-tuning.
We show that a core of the size currently observed 
is a natural consequence of two-body relaxation acting over
$10$ Gyr, starting from a core of radius $\sim 1-1.5$ pc;
the $n\sim r^{-2}$ density profile outside the core gradually
extends inward as the core shrinks and as the stellar density
evolves toward, but does not fully reach, the Bahcall-Wolf form after $10$ Gyr.

The relaxation time that sets the rate of evolution in these
models depends inversely on the mean stellar mass, roughly $1 \msun$
under standard assumptions about  the initial mass function
\citep{Alexander:05}.
But an old population also contains stellar remnants, including
$\sim 10\msun$ black  holes.
In the absence of a core, black holes initially at distances of 
$4-5$ pc from \sgr\ would spiral all the way in to the center
after $10$ Gyr \citep{Morris:93}, potentially dominating the
total mass density inside $\sim 10^{-2}$ parsec.
In the models considered here, the black hole orbits tend to decay no farther
than the core radius as determined by the stars (\S 5).
The result, after several Gyr, is a rather different distribution
of black holes than in the collisionally-relaxed models.
This difference has potentially important implications for the
rates of gravitational wave driven inspirals or for other dynamical
processes that postulate a dense cluster of black holes around
\sgr, as discussed in \S 6.

The observations motivating this paper are fairly new, and if recent
history is any guide, our observational understanding of 
the Galactic center will continue to  change.
It is conceivable that the low apparent density of late type stars is
an artifact due to  faulty stellar classifications, improper treatment of
extinction, confusion, or some other factor.
Stars slightly fainter than the current limit for robust detection
($m_K\approx 15.5$) may turn out, once detected, to have a very different
spatial distribution, more similar to that predicted by Bahcall and Wolf
(1976).
Our theoretical understanding of stellar collisions may also change;
for instance, new, more effective channels for removal of stellar
envelopes could be discovered.
Any of these developments would lessen the relevance of the models
discussed here to the center of the Milky Way.
However, core models are also applicable
to other galaxies that contain SMBHs, many of which have cores
that can not plausibly be explained except in terms of a general
depletion of stars.
And a robust conclusion to be drawn from this work is that
the distribution of stars and stellar remnants at the center of the
Milky Way should still reflect to some extent the details of the 
Galaxy's formation  -- though the imprint of the initial conditions
may turn out to be less extreme than in the models considered here.

We assume throughout a distance to \sgr\ of 8.0 kpc 
\citep{Eisenhauer:05,Gillessen:09}.
At this distance, one arc second corresponds to a linear distance
of $0.0388$ pc and one parsec corresponds to $25.78''$.
We also fix the SMBH mass to $4.0\times 10^6\msun$ 
\citep{Ghez:08,Gillessen:09}.

\section{Properties of the stellar distribution}
\subsection{Number density}

As summarized above, a relatively low density of late-type stars near the
center of the Milky Way was independently inferred by Buchholz et al. (2009),
Do et al. (2009) and Bartko et al. (2009).
The first of these three studies was based on the largest sample of
stars, and we adopt the Buchholz et al. number counts as basis for 
the discussion that follows.

\begin{figure}
\includegraphics[clip,width=0.45\textwidth]{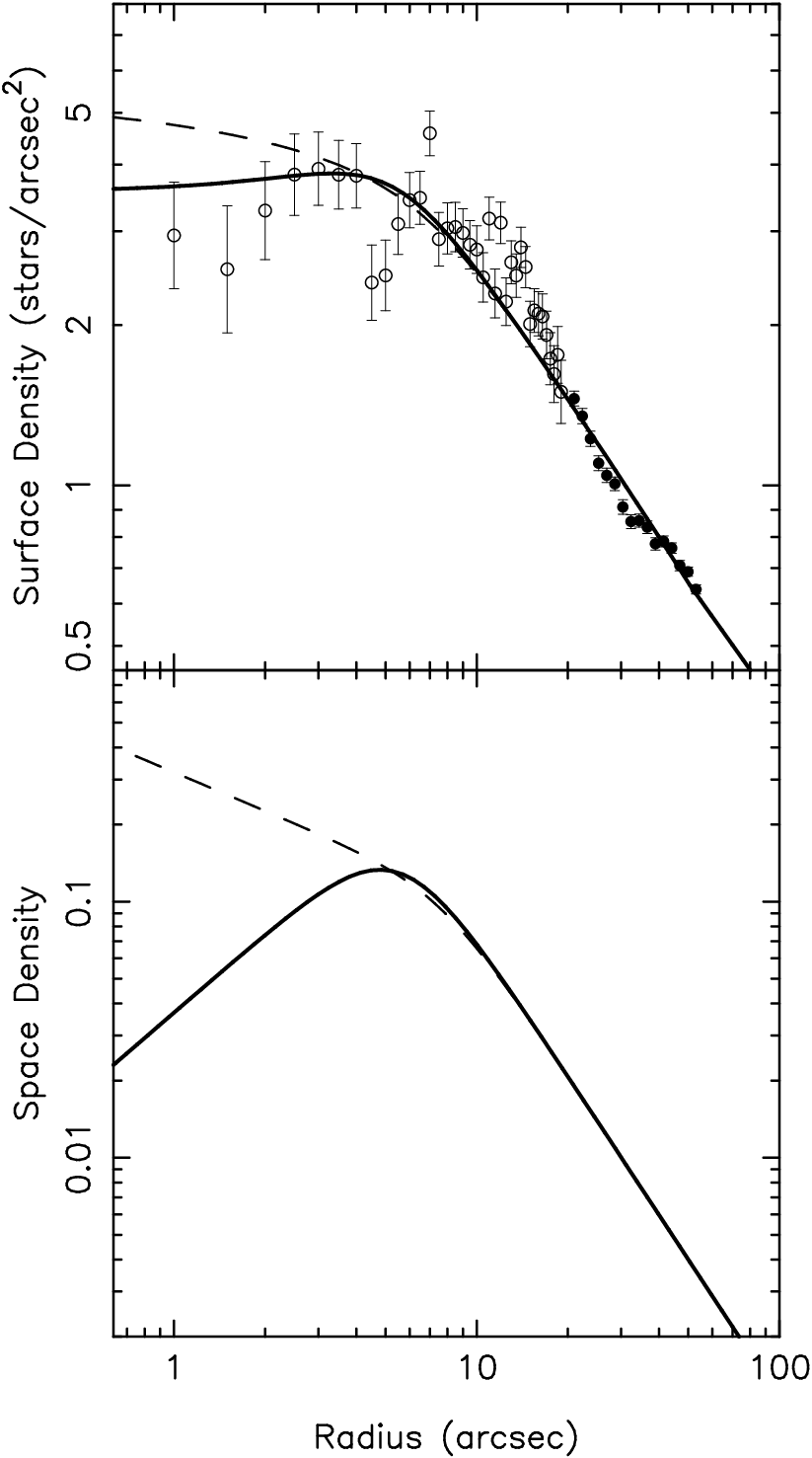}
\caption{{\it Upper panel:} azimuthally-averaged surface density of 
late type  stars at the Galactic center.
Open circles are 
number counts of the ``qualtity 1'' late type  stars
brighter than $K=15.5$ in  the sample of Buchholz et al. 
(2009) (their Fig. 11).
Filled circles are derived from a kernel estimate of
the projected density of all stars with
$K\le 15.5$ and $R\ge 20''$ in the sample of Sch\"odel et al. (2007),
after corrections for crowding and completeness.
Curves show fits of the projected,  parametric model, 
equation~(\ref{eq:nofr}), to the number count data.
Heavy curve is the best fit when the inner power-law slope is 
unconstrained;
dashed curve shows the best fit when the inner slope is set 
at $-0.5$, the shallowest
profile consistent with an isotropic velocity distribution.
Both fits assume $\alpha=4$ and $\gamma=1.8$.
{\it Lower panel:} space density profiles corresponding to the
projected profiles in the upper panel.
}
\label{fig:fits}
\end{figure}

Buchholz et al. (2009; hereafter BSE09) took deep, narrow-band images
($K<15.5$) in the near-IR $K$ band and distinguished late type (old) 
from early type
(young) stars using CO equivalent widths, calibrated from existing
samples.
Extinction was estimated star by star by comparing observed
SEDs with a blackbody of variable extinction.
BSE09 identified a ``quality 1'' sample
consisting of late type stars in which the CO band depth
exceeded the cutoff value for early type  stars by $1\sigma$ or more,
and which did not fall into any of the other classes defined by them
(AGB stars, foreground sources, very red objects).
The resulting data set contains 2955 stars down to a 
magnitude limit of $K=15.5$.
Figure~\ref{fig:fits} shows the BSE09 number counts (their Fig.~11)
plotted versus projected distance from \sgr.

The BSE09 data are believed complete to a limiting magnitude
$K=15.5$ within a projected distance of $\sim 20''$ from \sgr. 
Beyond this radius, the early type stars are insignificant in
numbers compared with the late type stars.
Sch\"odel et al. (2007) present total number counts in a more
extended region, derived from the ISAAC NIR camera on the VLT.
These counts extend to $\sim 60''$ and to a limiting magnitude of
$m_K\approx 17$.
Down to the magnitude limit ($K=15.5$) of the BSE09 sample, the ISAAC
counts are $\sim 80\%$ complete beyond $\sim 15''$.

Figure~\ref{fig:fits} includes number counts from all stars in
the ISAAC sample with $K\le 15.5$, and within the region 
$20''\le R\le 60''$.
The observed positions were converted to projected densities using the
kernel routine described in Sch\"odel et al. (2007) including
corrections for completeness and crowding, as discussed in that paper.
Confidence intervals were derived via the bootstrap.
The ISAAC counts are seen to match very well onto the BSE09 counts, 
without the necessity for any ad hoc adjustment of the normalizations.

At larger radii, out to $R \approx 10 {\rm pc}\approx 200''$,
a number of studies have found $n\sim r^{-1.8}$ for the distribution
of stellar light
\citep{BN:68,Haller:96,Genzel:00,Schoedel:07}.
In all of what follows, we assume an asymptotic slope 
$d\log n/d\log r=-1.8$.

This leads us to parametrize the space density
of the old stellar population as a power law with an inner core:
\beq
n(r) = n_0 \left({r\over r_0}\right)^{-\gamma_i}
\left[1+\left({r\over r_0}\right)^\alpha\right]^{(\gamma_i - \gamma)/\alpha}
\label{eq:nofr}
\eeq
with $\gamma=1.8$.
The parameter $\alpha$ controls the sharpness of the transition
from outer to inner slopes.
This model was fit to the number counts after projecting it
onto the plane of the sky;
in this way one avoids fitting a model to the surface density
that does not correspond to a physical space density.

As is often found when modelling the luminosity profiles of
cored elliptical galaxies \cite[e.g.][]{Terzic:05},
large values of $\alpha$ gave the best fits.
The thick curve in Figure~\ref{fig:fits} shows the best-fit
model when $\alpha=4$; it has
\beq
n_0 = 0.21 {\rm pc}^{-3}, \ \ r_0=5.5'' = 0.21\ {\rm pc}, \ \ \gamma_i = -1.0.
\label{eq:fit1}
\eeq
The negative value of $\gamma_i$ implies a space density
that {\it decreases} toward the center
(lower panel of Fig.~\ref{fig:fits}) -- a central ``hole.''

Fits of this model to the data were found never to be terribly good
($\tilde\chi^2\gap 17$).
There is a broad local maximum in the number counts at $R\approx 15''$
which would require additional parameters to fit.
(In the 2d counts, Figure~13 of BSE09, this overdensity
is seen to be roughly symmetric about the origin.)
The central minimum in the projected density is also difficult
to reproduce.

\begin{figure}
\includegraphics[clip,width=0.45\textwidth]{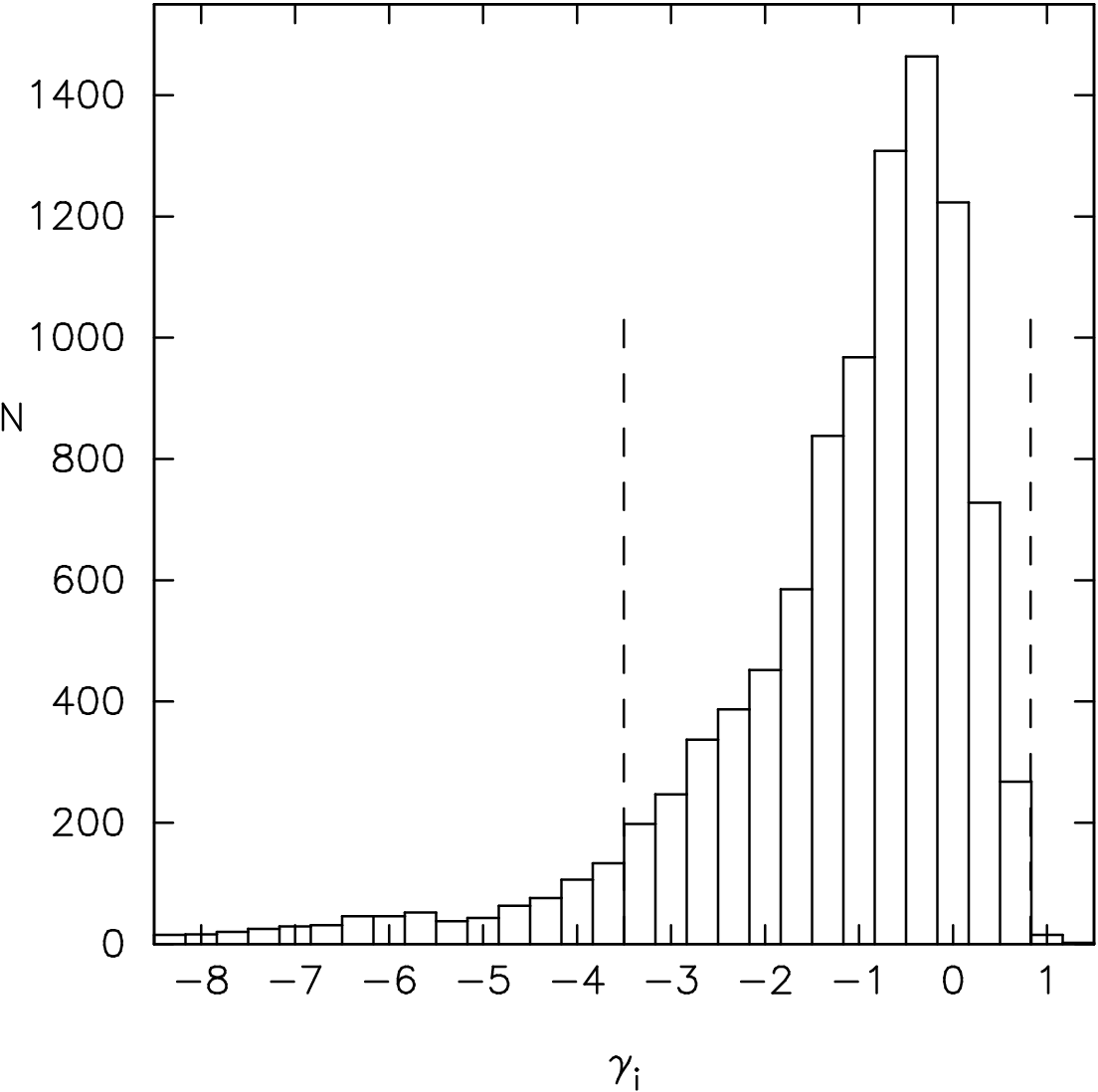}
\caption{Distribution of inner slopes from fits to $10^4$ Monte-Carlo
samples bootstrapped from the number count data in Fig.~\ref{fig:fits}.
Dashed lines delineate a 90\% confidence interval.
}
\label{fig:mc}
\end{figure}

Figure~\ref{fig:mc} shows the distribution of inner slopes
$\gamma_i$ obtained from the fits to $10^4$ Monte-Carlo
samples bootstrapped from the number count data in Figure~\ref{fig:fits}.
90\% of the values fall in the range
\beq
-3.5 \le \gamma_i \le 0.82.
\eeq
Negative values of $\gamma_i$ (corresponding to centrally-decreasing
densities) are clearly preferred, though positive values 
are also acceptable.

If the stellar velocity distribution is isotropic,
the space density {\it must} increase at least as fast
as $r^{-0.5}$ near the SMBH (\S 3).
The dotted curve in Figure~\ref{fig:fits} shows the best fit
when $\gamma_i$ is fixed at $1/2$.
The other parameters are
\beq
n_0 = 0.12 {\rm pc}^{-3}, \ \ r_0=7.7'' = 0.30\ {\rm pc}.
\label{eq:fit2}
\eeq
This model can be (crudely) thought of having the minimum central density 
consistent with an isotropic velocity distribution.
This fit is only slightly worse in a $\chi^2$ sense ($19.0$ vs. $17.5$)
than the fit with unconstrained inner slope, although it implies
a very different {\it space} density profile inside $\sim 0.5$ pc
(Fig.~\ref{fig:fits}, lower panel).

A standard definition of the core radius is the projected radius 
where the surface density falls to $1/2$ of its central value 
\cite[e.g.][]{King:62}.
As a practical definition of the ``central density,'' 
we take the value at $1''\approx 0.04$ pc projected radius.
Based on this definition, the core radius of the unconstrained
fit in Figure~\ref{fig:fits} is
\beq
\rc\approx 15.2 '' \approx 0.59\ {\rm pc}
\eeq
and the core radius of the constrained fit
($\gamma_i=-1/2$) is
\beq
\rc\approx 10.8 '' \approx 0.42\ {\rm pc}.
\eeq
Thus, the core radius of the Milky Way nuclear star cluster
is $\sim 0.5$ pc.

\subsection{Relaxation time}

The relatively low central density of stars implied by  
Figure~\ref{fig:fits}, and the apparently
non-relaxed form of $n(r)$, are suggestive of a long
two-body relaxation time.
Spitzer (1987) defines the local relaxation time as
\begin{eqnarray}
t_r &=& {0.33\sigma^3\over G^2n m^2\ln\Lambda} \\
&=& 1.2\times 10^{9}{\rm yr} {\left[\sigma({\rm km/s})\right]^3\over
\rho(\msun{\rm pc}^{-3})\left[m/\msun\right] \left[\ln\Lambda/15\right]}
\label{eq:tr_spitz}
\end{eqnarray}
where $\sigma$ is the rms velocity in any direction,
$\rho$ is the stellar mass density,
$m$ is the mass of one star, and $\ln\Lambda$ is the Coulomb logarithm.
This expression assumes that all stars have the same mass
and that their velocity distribution is isotropic (Maxwellian).

To apply equation~(\ref{eq:tr_spitz}) we need an estimate of $\rho(r)$.
Dynamical estimates of $\rho$ are to be preferred, given the large
systematic uncertainties associated with converting a luminosity
density into a mass density near the Galactic center 
\citep{Schoedel:07,BSE:09}.

Inside $\sim 1$ pc, the best dynamical constraints on $\rho$
come from the recent proper motion study of Sch\"odel et al. (2009).
These authors detected, for the first time, an unambiguous
signature of the gravitational force from the distributed mass on the 
stellar motions, in the region $0.25\lap r\lap 1$ pc; 
inside this region the gravitational force from the SMBH can not be 
disentagled from that of the stars.
Sch\"odel et al. inferred a distributed mass of 
$\sim 1 \pm 0.5 \times 10^6\msun$ in a sphere of radius $1$ pc around \sgr.
The functional form of $\rho(r)$ was not well constrained, although
the formal best fits were obtained with models in which the mass
density decreased toward the center.

\begin{figure}
\includegraphics[clip,width=0.45\textwidth]{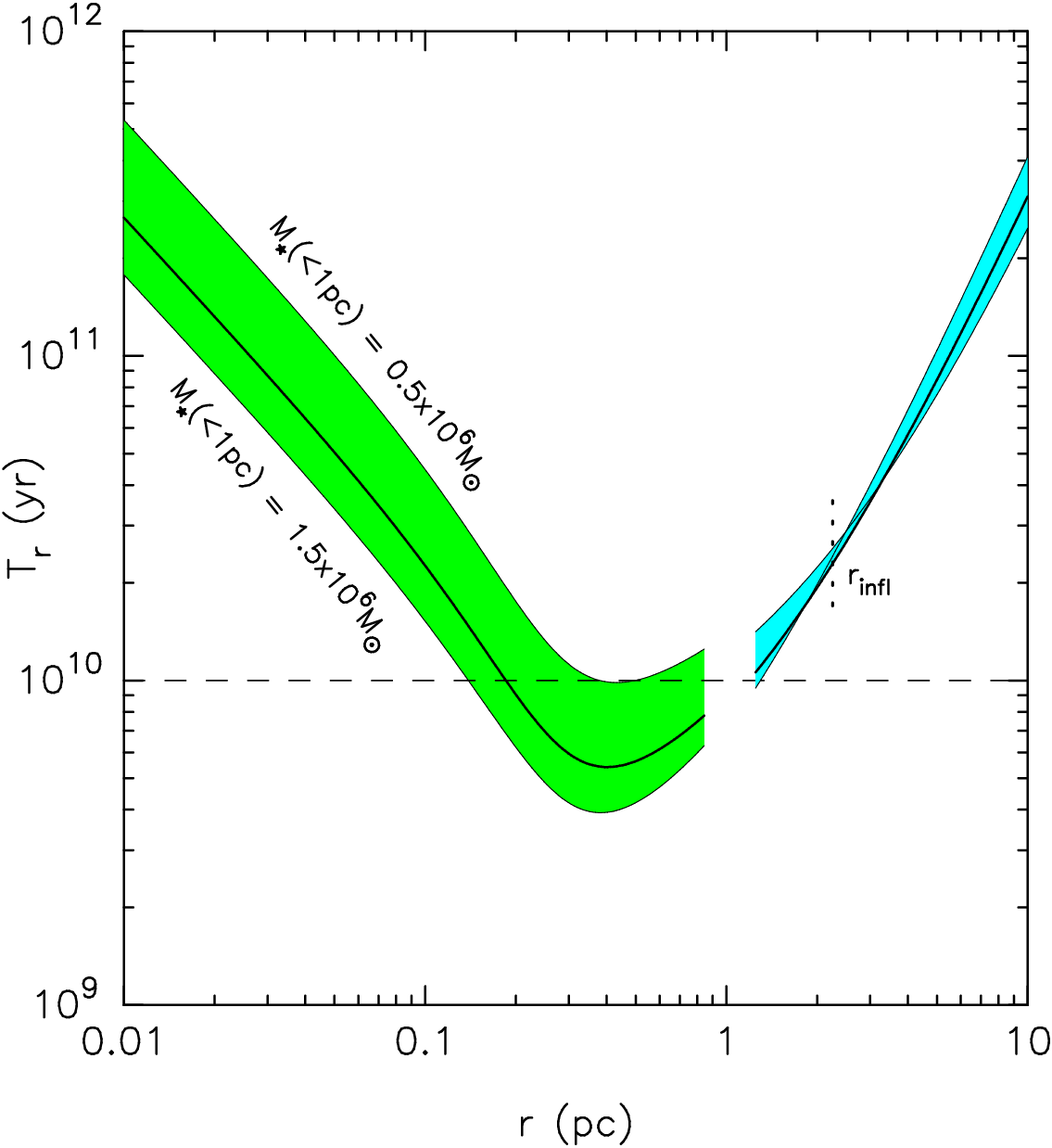}
\caption{Relaxation time vs. radius at the Galactic center.
Curves on the left (green filled region) 
assume the mass density of eq.~(\ref{eq:rho2}) with three
normalizations, such that the total (distributed) mass within 1 pc is 
$\tilde M_\star\equiv M_\star(<1 {\rm pc})/\msun = 
(0.5,1.0,1.5) \times 10^6$; the $\tilde M_\star=1.0\times 10^6$ model
is shown as the heavy line.
Curves on the right (blue filled region) assume
$\rho\propto r^{-1.8}$ at all radii with various values for 
the normalizing density at one parsec,
$\rho_0 = (0.75,1.5,3)\times 10^5 \msun {\rm pc}^{-3}$;
the $\rho_0 = 1.5\times 10^5 \msun {\rm pc}^{-3}$ (preferred)
model is shown with the heavy line.
Vertical tick mark is the SMBH influence radius
computed using $\sigma(r)$ from the preferred model.
}
\label{fig:tr}
\end{figure}

We use the Sch\"odel et al. results to normalize the density
of our ``isotropic'' $n(r)$, equations~(\ref{eq:nofr}, \ref{eq:fit2}).
Defining
\beq
\tilde M_\star \equiv {M_\star(r\le 1 {\rm pc})\over 10^6\msun} \approx 1,
\eeq
the mass density becomes
\beq
\rho(r) = 9.9\times 10^5 \msun {\rm pc}^{-3}\tilde M_\star
\xi^{-0.5} \left(1+\xi^4\right)^{-0.325}
\label{eq:rho2}
\eeq
with $\xi=r/0.30$ pc.
We then calculate the (isotropic) velocity dispersion from the Jeans equation,
\beq
\rho(r)\sigma(r)^2 = G\int_r^{\infty} dr' r'^{-2} \left[\mh+M_\star(<r')\right]\rho(r').
\label{eq:jeans}
\eeq
Figure~\ref{fig:tr} shows the resulting $t_r(r)$ assuming 
$\ln\Lambda=15$, $m=\msun$ and three values of $\tilde M_\star$.

The Sch\"odel et al. (2009) proper motion data do not constrain
the density beyond $\sim 1$ pc.
In this region, we assume
\beq
\rho(r) = 10^5\tilde\rho  \left({r\over 1 {\rm pc}}\right)^{-1.8} \msun {\rm pc}^{-3},
\label{eq:rho}
\eeq
a simple continuation of  the model fit at smaller radii.
Below we argue for a ``preferred'' density model with $\tilde\rho=1.5$,
but in Figure~\ref{fig:tr} we allow the normalization:
\beq
\tilde\rho \equiv {\rho(1 {\rm pc})\over 10^5\msun} 
\eeq
to have the values $\tilde\rho = (0.75,1.5,3)$.
The Jeans equation was again used to compute $\sigma(r)$.
The resulting $t_r(r)$ curves match well onto the curves at smaller radii
and there is little dependence of $t_r$ on $\tilde\rho$ within this 
radial range.

Also shown in Figure~\ref{fig:tr} is the influence radius
$\rh$ of the SMBH, defined as the root of the equation
\beq
\sigma^2(x) = {G\mh\over x}.
\eeq
Using the preferred density model with $\tilde\rho=1.5$, 
one finds $\rh\approx 2.5$ pc.
The relaxation time at $r\approx \rh$ is a reasonable estimate of the time
scale over which gravitational encounters can change the gross
properties of the core.
Figure~\ref{fig:tr} shows that the relaxation time at $\rh$ is 
$\gap 2-3\times 10^{10}$ yr.

The estimates of $t_r$ in Figure~\ref{fig:tr} assume 
a single population of $1\msun$ stars.
In reality the Galactic center contains a range of mass groups,
each of which might have a different velocity distribution.
Given the relatively long relaxation time,
a natural case to consider is a distribution of stellar masses,
all of which have the same velocity dispersion at every radius.
Equation~(\ref{eq:tr_spitz}) becomes in this case
\begin{subequations}
\begin{eqnarray}
t_r &=& {0.33\sigma^3\over \rho \tilde m G\ln\Lambda}, 
\label{eq:tr_spitz2}\\
\tilde m &=& {\int N(m) m^2 dm\over \int N(m) m dm}
\label{eq:defmtilde}
\end{eqnarray}
\label{eq:both}
\end{subequations}
and $N(m)dm$ is the number of stars with masses in the range
$m$ to $m+dm$ \cite[e.g.][]{Merritt:04}.
Since equation~(\ref{eq:tr_spitz}) was derived from the 
diffusion coefficient $\langle(\Delta v_\|)^2\rangle$
describing gravitational scattering,
equation~(\ref{eq:tr_spitz2}) is properly  interpreted as the 
time for a test star's velocity to be randomized by encounters with
more massive objects.
Making standard assumptions about the initial mass function
gives $\tilde m\approx 1\msun$ \citep{Merritt:04};
if the density is dominated locally by stellar black holes
then $\tilde m\approx\mbh\approx 10\msun$.
It has been argued that $\tilde m$ may be even
larger outside the central few parsecs due to giant
molecular clouds  \citep{Perets:07}.

\section{Making a core}
\label{sec:core}

The classical model for a core \cite[e.g.][]{Tremaine:97} is
a region of constant gravitational potential $\phi$ and constant
phase-space density $f$ near the center of a galaxy.
If the central potential is dominated by a SMBH, 
$\phi(r) \approx -G\mh/r$ and a constant
$f$ translates into a steeply-rising $\rho$:
\beq
\rho = \int fd^3\mathbf{v} \propto f\int _0^{\sqrt{-2\phi(r)}} v^2 dv 
\propto f (-2\phi)^{3/2} \propto r^{-3/2}
\label{eq:adiabat}
\eeq
inside $\sim\rh$ \citep{Peebles:72}, inconsistent with the observed
distribution.

The Galactic center has a core of size $\sim 0.5$ pc, smaller than
$\rh\approx 2.5$ pc (Figs.~\ref{fig:fits},~\ref{fig:tr}) 
so equation~(\ref{eq:adiabat}) would apply in this region.
Making a core similar to the observed core therefore requires a reduction 
in the value of $f$ on orbits that pass inside $\sim 0.5$ pc.

There are of course many ways to do this.
Given that the number counts do not strongly constrain
the form of $n(r)$ within the core (Fig.~\ref{fig:mc}), 
we choose not to solve
the inverse problem $n\rightarrow f$.
Instead, we focus on two simple core models,
both motivated by physical arguments, that are consistent with the
number count data.

The starting point is a power-law density of 
stars at all radii around the SMBH,
$\rho_i\propto r^{-\gamma}$, $0\le r\le\infty$.
For definiteness, we normalize the density of this initial
(i.e. coreless) model to be
\beq
\rho_i(r) = 1.5\times 10^5 \left({r\over 1 {\rm pc}}\right)^{-1.8} 
\msun {\rm pc}^{-3}
\label{eq:rhoi}
\eeq
and we assume $n_i(r)\propto\rho_i(r)$.
(The latter assumption will be relaxed in \S 5.)
Equation~(\ref{eq:rhoi}) is just equation~(\ref{eq:rho}) with $\tilde\rho=1.5$.
The normalizing constant is similar to what various authors have
derived in the past for the mass density at $1$ pc; 
e.g. $\tilde\rho\approx 1.8$ \citep{Genzel:03},
$\tilde\rho\approx 2.0$ \citep{Schoedel:07} etc.
The exact normalization is not critical in what follows:
it serves mostly to fix the relaxation time, and
as Figure~\ref{fig:tr} shows, $t_r$ in the region of interest
is weakly dependent on the density normalization.
We show below that our ``preferred'' value of $\tilde\rho$ gives a good
fit to the observed stellar velocities at $r\approx\rh$.

The stellar mass implied by this model inside one parsec is 
$\sim 1.6\times 10^6\msun$.
This is the mass {\it before} the core has been carved out; for consistency,
this mass should exceed the dynamically inferred (distributed) mass inside
one parsec, $\sim 1\pm 0.5 \times 10^6\msun$  \citep{Schoedel:09},
and it does.
At the same time, the proper motion data are consistent with
the mass implied by the unmodified power law model, so we can
not robustly infer the presence of a core in the Galactic center {\it mass} 
distribution from the proper motion data alone.

The gravitational potential generated by $\rho_i(r)$, 
including the contribution from the SMBH, can be written
\begin{eqnarray}
\phi(r) - \phi(r_0) &=& {G\mh\over r_0} \left(1-{r_o\over r}\right) \\
&&- {1\over 2-\gamma} {GM_0\over r_0} 
\left[1-\left({r\over r_0}\right)^{2-\gamma}\right]
\end{eqnarray}
where $r_0$ is a fiducial radius, taken in what follows to be $1$ pc, 
$M_0=10^6\tilde M_\star\msun$ is the distributed mass inside $r_0$, 
and $\gamma=1.8$.
The isotropic distribution function $f_i(E)$ corresponding to the
pair of functions ($\rho_i,\phi$) can be derived numerically
from Eddington's (1916) formula:
\beq
f_i(E) = {1\over m}{1\over \sqrt{8}\pi^2} {d\over dE} \int_{\phi(r)}^E
{d\rho_i\over d\phi}
{d\phi\over\sqrt{E-\phi}}
\label{eq:fi}
\eeq
where $E=v^2/2 + \phi(r)$ is the energy per unit mass of a star.
(Note that the subscript $i$ refers here to ``initial,'' not
``isotropic.'')
Isotropy is a reasonable assumption, although as we will show,
some reasonable models for a core imply substantial anisotropy.

\subsection{Core origins}

Before proceeding, we consider possible mechanisms for the formation
of a parsec-scale core at the center of a galaxy like the Milky Way.

1. {\it A binary supermassive black hole.}
In giant elliptical galaxies, cores are often attributed to ejection
of stars by a pre-existing binary SMBH \citep{Faber:97,MM:01},
and possibly to gravitational-wave recoil after binary coalescence
\citep{BMQ:04,GM:08}.
This model naturally explains the sizes of many observed cores -- 
comparable to the influence radius of the (single) observed SMBH -- 
if it is assumed that the galaxy grew through at least one ``major merger,'' 
with comparably-massive SMBHs \citep{Merritt:06a}.
However the binary SMBH model seems less relevant to a disk-dominated
system like the Milky Way, which may never have experienced a major merger.
Based on a standard $\Lambda$CDM cosmological model,
the probability that the Milky Way has avoided accreting a galaxy with
halo mass $1/4$ that of the Milky Way or greater, since a redshift of $z=2$,
is $\sim 30$\% \citep{MMVJ:02}.
The most recent major merger is likely to have occurred 10-12 Gyr ago,
around the time of formation of the thick disk \citep{Wyse:01}.

2. {\it  Inspiral of multiple, smaller black holes}.
A single intermediate-mass black hole (IMBH) of mass $\sim 10^4M_\odot$,  
spiralling in against a pre-existing stellar density cusp,
creates a core of radius $\sim 0.05-0.1$ pc \citep{BGZ:06}.
Repeated inspiral events could create a larger core,
although the displaced mass would increase at a less than linear
rate with the number of inspirals.
Nevertheless, some models postulate one such event
every $\sim 10^7$ yr \citep{Zwart:06};
if so, more than one IMBH would probably be 
present at any given time in the inner parsec.

3.  {\it An enlarged loss cone.} 
Gravitational encounters drive a mass flux 
of $\sim \mh/t_r(r_\mathrm{infl})$ into Sgr A$^*$.
The core that results from this diffusive loss process is very
small: its size is comparable to the radius of the capture sphere --
either the tidal disruption radius,
$r_t\approx 10^{-5}$ pc, or the Schwarzschild radius,
$r_\mathrm{Sch}\approx 10^{-6}$ pc.
The core is small because the depleted orbits are
continuously resupplied by diffusion from
orbits of larger angular momentum and energy.
If there were some way to transfer a mass in stars of $\sim \mh$
into the SMBH on a time scale $\ll t_r$ -- say, a crossing time --
the resulting core would be much larger.
This could happen if the NSC were appreciably triaxial,
even if only transiently, since many orbits near a SMBH in a triaxial cluster 
are ``centrophilic,''
passing arbitrarily close to the SMBH after a finite time
(some multiple of the crossing time) \citep{MP:01}.
The size of the resultant core is determined by a number of factors,
including the degree of non-axisymmetry and the population of the
various orbit families, but it could be of order $\sim \rh$ \citep{MP:04}.

4. {\it Localized star formation.} Stars might form only, or preferentially, 
beyond a certain radius from SgrA$^*$, resulting in a low density
inside this radius.
This possibility is discussed in more detail in \S\ref{sec:discussion}.1.

5. {\it Feedback in active nuclei} has also been proposed as
a rapid core-formation mechanism \citep{Peirani:08}.

\begin{figure*}
\includegraphics[clip,width=0.425\textwidth,angle=-90.]{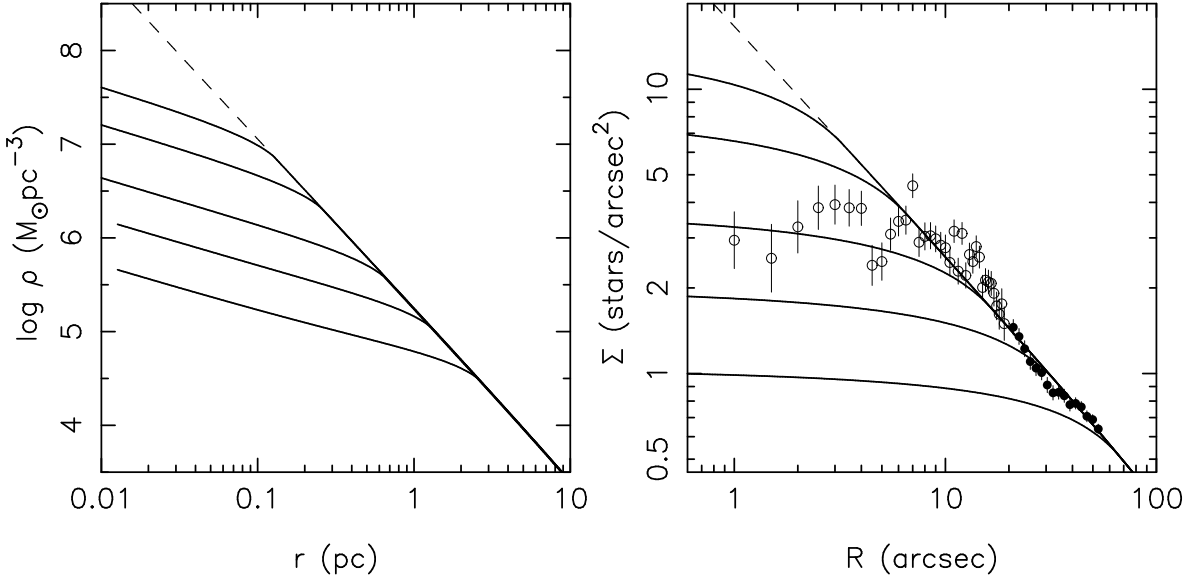}
\caption{``Isotropic'' core models, created by setting $f=0$ at low
(bound) energies, $E\le E_b\equiv \phi(r_b)$, starting from a power-law
model in the density, $\rho\propto r^{-1.8}$ (shown as the dashed line
in both panels).
Left panel shows space densities for $r_b=(0.1,0.2,0.5,1,2)$ pc.
Right panel shows surface densities of the same models, compared
to  the number-count data from Figure~\ref{fig:fits}.
The vertical normalization of the models in this panel was chosen
arbitrarily .
}
\label{fig:core1}
\end{figure*}

\subsection{Isotropic core}

A simple way to lower the density of stars near the SMBH
is to set 
\beq
f(E)=0,\ \ \ \  E\le E_b.
\eeq
Such a truncation leaves the velocity distribution isotropic.
The cutoff energy can be expressed in terms of a cutoff radius $r_b$ where
\beq
E_b \approx \phi(r_b).
\label{eq:defeb}
\eeq
The configuration-space density after removal of the most-bound
stars is
\begin{subequations}
\begin{eqnarray}
\rho(r) &\approx& \rho_i(r), \ \ r \ge r_b \\
&\approx& 4\sqrt{2}\pi\int_{\phi(r_b)}^\infty 
dE f_i(E) \sqrt{\phi(r)-E} , \ \ r < r_b.
\end{eqnarray}
\label{eq:rhocore1}
\end{subequations}
The ``approximately equal'' sign in these expressions is due to the
fact that removal of the most-bound stars causes a change in the
gravitational potential.
We ignore that complication in what follows, i.e. we replace
$\approx$ by $=$ in equations~(\ref{eq:defeb}),
(\ref{eq:rhocore1}) and similar expressions below.
The resultant error in $\rho$ is at most a few per cent, since $\phi$
is dominated by the SMBH at $r\lap\rh$, and by the stellar
potential due for $r\gap\rh$.

At small radii, $r\ll r_b \lap \rh$, the density is
\beq
\rho \approx 4\sqrt{2}\pi\sqrt{-\phi(r)}
\int_{\phi(r_b)}^\infty dE f_i(E) 
\propto r^{-1/2}.
\label{eq:rhalf}
\eeq
In spite of the zero-density hole in phase space,
the configuration-space density 
diverges, mildly, toward the SMBH.
This demonstrates that an isotropic $f$ is not consistent with
a strictly flat core, much less with
a central dip in $\rho$.

Figure~\ref{fig:core1} shows space and projected density profiles
for various values of $r_b$.
The observed number counts are reasonably well fit by a model
with
\beq
r_b \approx 0.5\ {\rm pc}.
\eeq
This value of $r_b$ is close to the core radius derived above from 
the number counts.
In other words, in these simple core models, $r_b\approx r_{\rm core}$.

\begin{figure*}
\includegraphics[clip,width=0.425\textwidth,angle=-90.]{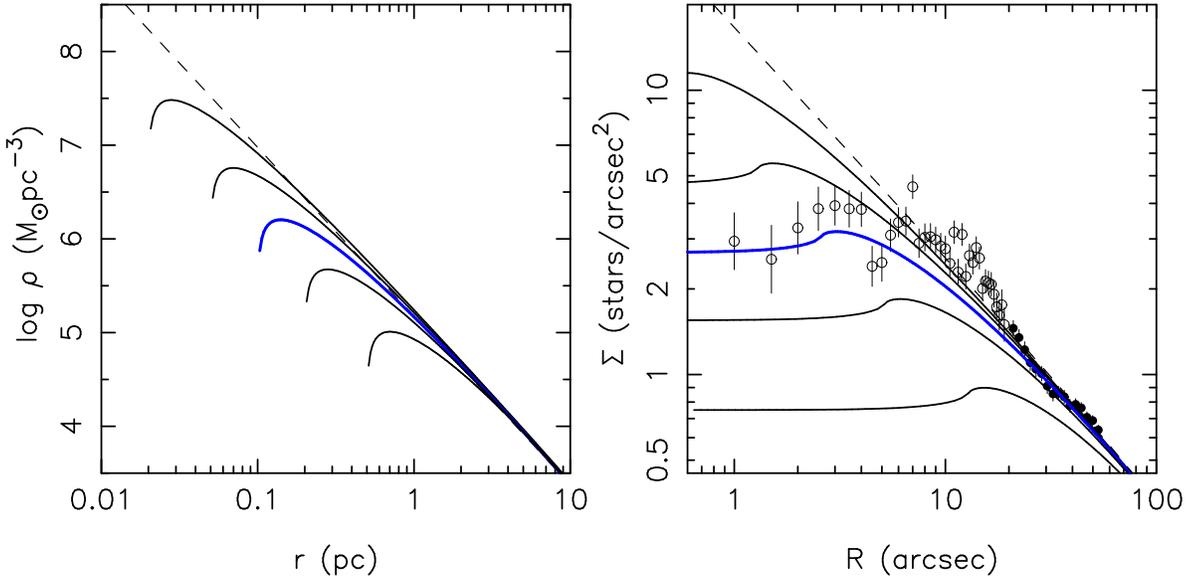}
\caption{``Ansotropic'' core models, created by setting $f=0$ on orbits
that pass below $r_b$.
Left panel shows space densities for $r_b=(0.02,0.05,0.1,0.2,0.5)$ pc.
Right panel shows surface densities of the same models, compared
to  the number-count data from Figure~\ref{fig:fits}.
Blue (thick) curve is marked for comparison with the same
model in Figure~\ref{fig:anisot}.
The vertical normalization of the models in the right panel was chosen
arbitrarily.
}
\label{fig:core2}
\end{figure*}
\begin{figure*}
\includegraphics[clip,width=0.425\textwidth,angle=-90.]{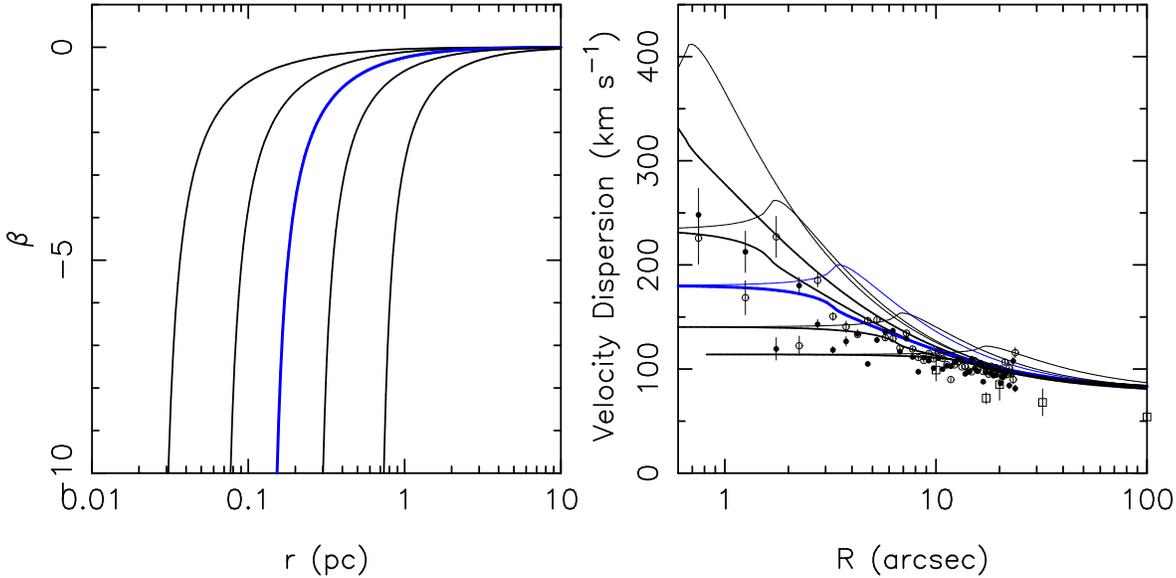}
\caption{Kinematics of the models of Figure~\ref{fig:core2}.
{\it Left panel:} anisotropy profiles.
{\it Right panel:} projected, radial (thick lines) and tangential 
(thin lines) velocity dispersions
in the plane of the sky, compared with the observed
$\sigma_R$ (filled circles) and $\sigma_T$ (open circles)
from Schoedel et al. (2009).}
\label{fig:anisot}
\end{figure*}

The  strange character of these (idealized)  models -- a zero
phase space density and a non-zero configuration space density --
implies some other strange properties.
In the Appendix, the distribution of orbital elements for stars
passing near the SMBH is derived.
In spite of the isotropic $f$, the orbits near the center are
very eccentric, with an expected size $a\approx 0.2r_b\approx 0.1$ pc.

\subsection{Anisotropic  core}

Figure~\ref{fig:fits} suggests that $n(r)$ may be decreasing
toward \sgr.
As our second model, we create a core by removing
all stars on orbits that pass within the sphere $r=r_b$ around
the SMBH.
Such a model is a crude description of what happens when
a binary SMBH ejects stars:
in this case, $r_b$ would be approximately the binary semi-major axis.
Collisional destruction of (all) stars that pass within a distance
$r_b$ of the SMBH would also result in such a truncation of $f$.

Since we are preferentially removing stars on low-angular-momentum orbits,
the velocity distribution in the core will be anisotropic, biased toward 
circular motions.

We define $E_b=\phi(r_b)$ as before, and 
$E_c$ is the energy (kinetic plus potential)
of a test particle on a circular orbit of radius $r_b$.
If the core is small, $r_b\ll\rh$, then $E_c\approx E_b/2$.
Starting again from a power-law density profile, equation~(\ref{eq:rhoi}),
and an isotropic distribution function, equation~(\ref{eq:fi}), 
we now set $f=0$ when either
\begin{subequations}
\begin{eqnarray}
E &\le& E_c \\
{\rm or} \ \ \ \ \ \ \ \ \ \ && \nonumber \\
J^2 &\le& J_b^2=2r_b^2\left(E-E_b\right);
\end{eqnarray}
\end{subequations}
$J_b(E)$ is the specific angular momentum of an orbit with energy $E$
and periapse $r_b$.
The configuration-space density is now zero inside $r_b$:
\beq
\rho(r) = 0, \ \ r\le r_b.
\eeq
Outside $r_b$, the density is lowered at {\it all} radii due to the
absence of low-$J$ stars:
\begin{eqnarray}
\rho(r) &=& {2\pi\over r^2} \int_{E_0(r)} dE f_i(E) 
\int_{J_b^2}^{2r^2(E-\phi(r))} {dJ^2\over 
\sqrt{2\left[E-\phi(r)\right] - J^2/r^2}} \nonumber \\
&=& 4\sqrt{2}\pi
\left(1 - {r_b^2\over r^2}\right)^{1/2} 
\int_{E_0(r)}^\infty dE f_i(E) \sqrt{E-E_0} .
\end{eqnarray}
Here
\beq
E_0 = {r^2\phi(r) - r_b^2E_b\over r^2 - r_b^2}
\eeq
is the minimum energy of orbits that pass through $r$ without
also passing below $r_b$.

Figure~\ref{fig:core2} shows density profiles for anisotropic core
models with various $r_b$.
The observed number counts are reasonably well fit by a model
with
\beq
r_b \approx 0.1 {\rm pc}.
\eeq
Note that a given value of $r_b$ produces a larger core
than in the isotropic core models, since a larger
region of phase space is affected.

Defining $\sigma_r$ and $\sigma_t$ as the
1d velocity dispersions in the radial and tangential directions,
one finds
\begin{subequations}
\begin{eqnarray}
&&\rho\sigma_r^2 = {8\sqrt{2}\pi\over 3} \left(1-{r_b^2\over r^2}\right)^{3/2}
\int_{E_0}^\infty dE f_i(E) \left(E-E_0\right)^{3/2}, \\
&&\rho\sigma_t^2 = \rho\sigma_r^2 + \nonumber \\
&& 4\sqrt{2}\pi {r_b^2\over r^2} \sqrt{1-{r_b^2\over r^2}} 
\int_{E_0}^\infty dE f_i(E)\sqrt{E-E_0} \left(E-E_b\right)
\end{eqnarray}
\end{subequations}
for $r>r_b$; at smaller radii $\rho=0$.
The anisotropy, defined in the usual way as
$\beta=1 - \sigma_t^2/\sigma_r^2$, is then
\beq
\beta(r) = -{3\over 2} {r_b^2\over r^2-r_b^2} 
{\int_{E_0}^\infty dE f_i(E) \left(E-E_b\right)\sqrt{E-E_0} \over
\int_{E_0}^\infty dE f_i(E) \left(E-E_0\right)^{3/2}}
\eeq
and is manifestly negative, i.e. $\sigma_t>\sigma_r$.

Figure~\ref{fig:anisot} plots anisotropy profiles for the models
in Figure~\ref{fig:core2} as well as projected velocity dispersion
profiles in the plane of the sky, compared with the observed,
proper-motion-based velocity dispersions from Sch\"odel et al. (2009).
The same values of $r_b$ that give a good fit to the number
count data, also appear to fit the proper motion data reasonably
well.
The tangential anisotropies predicted by these value of
$r_b$ -- which peak between $\sim 2$ and $\sim 5$ arc sec --
are consistent with what is observed, although the statistical
error bars are so large  that no clear discrimination between
theoretical models can be made.

This figure also shows that the adopted density normalization,
equation~(\ref{eq:rhoi}), yields velocities that are consistent with the 
proper motion data.

\section{Evolutionary models}

The distributions of stars illustrated in Figure~\ref{fig:fits},
and in the simple phase-space models of Figures~\ref{fig:core1} and
\ref{fig:core2}, are different from the collisionally relaxed
distributions normally associated with stars around a SMBH \citep{BW:76,BW:77}.
This may be a consequence of the relatively long relaxation time
at the Galactic center (Figure~\ref{fig:tr}).
A key question is whether models with a core can survive on
Gyr time scales.
In order to investigate this question, in this section we
consider time-dependent models for the phase-space
distribution, starting from initial conditions like those discussed
in \S\ref{sec:core}.

Models that are initially isotropic will remain close to isotropy as they
evolve.
Given such initial conditions, the isotropic, orbit-averaged Fokker-Planck
equation \citep{Henon:61} should provide a good description of the evolution.

Initially anisotropic models will evolve both in $J$- and $E$-space.
Here, we make the approximation that the evolution 
can be divided into two sequential phases: ``fast'' evolution in
$J$, on a time scale $\ll t_r$, followed by ``slow'' evolution
in $E$, on a timescale $\sim t_r$.
The justification follows from an argument made originally by
Frank \& Rees (1976):
Diffusion of stars into a point-mass ``sink'' is dominated by scattering
of low-$J$ orbits, on a time scale
\beq
t_\theta\approx \theta(r)^2t_r(r)
\eeq
where $\theta\approx (r_b/r)^{1/2}$ 
is the angle within which a star's velocity vector must
lie in order for its orbit to intersect the central sink -- in our
case, the edge of the low-density core at $r\approx r_b$.
The square-root dependence of $\theta$ on $t_\theta$ reflects the
fact that evolution in $J$ is a diffusive process.
Thus
\beq
t_\theta \approx {r_b\over r} t_r
\eeq
implying the separation of time scales for stars at $r\gap r_b$.

\begin{figure*}
\includegraphics[clip,width=0.425\textwidth,angle=-90.]{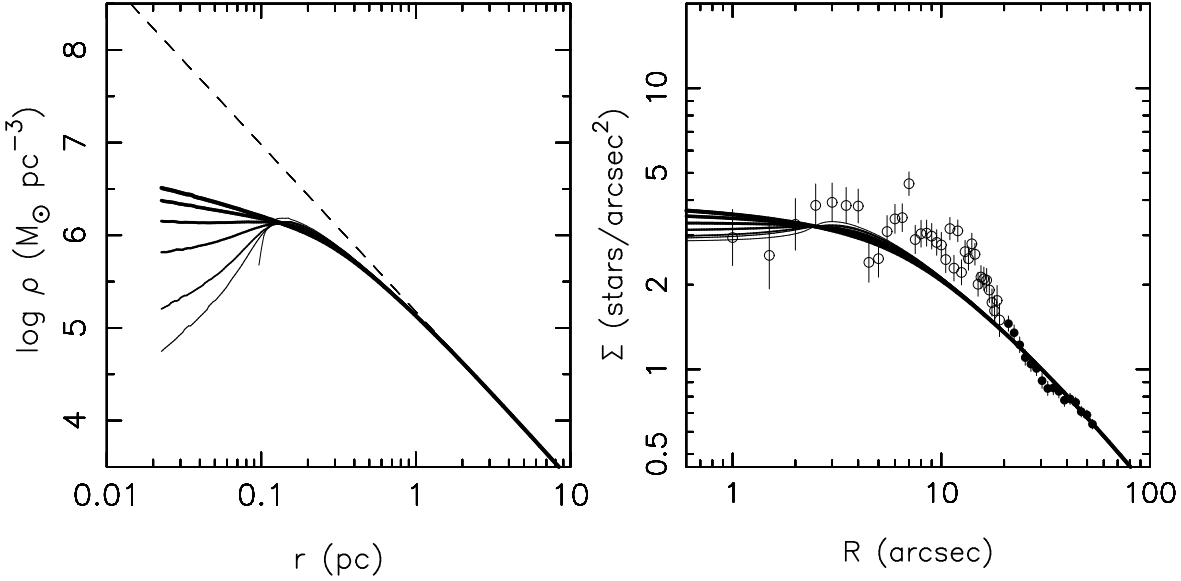}
\caption{Evolution of the density of the 
``anisotropic core'' model with $r_b=0.1$ pc
due to diffusion in $J$.
Curves show space (left) and projected (right)  densities at times
$(0,0.1,0.2,0.5,1,2,5)\times 10^9$ yr.
Line thickness increases  with time.
Other symbols are as in Figure~\ref{fig:core2}.
}
\label{fig:rhoj}
\end{figure*}
\begin{figure*}
\includegraphics[clip,width=0.425\textwidth,angle=-90.]{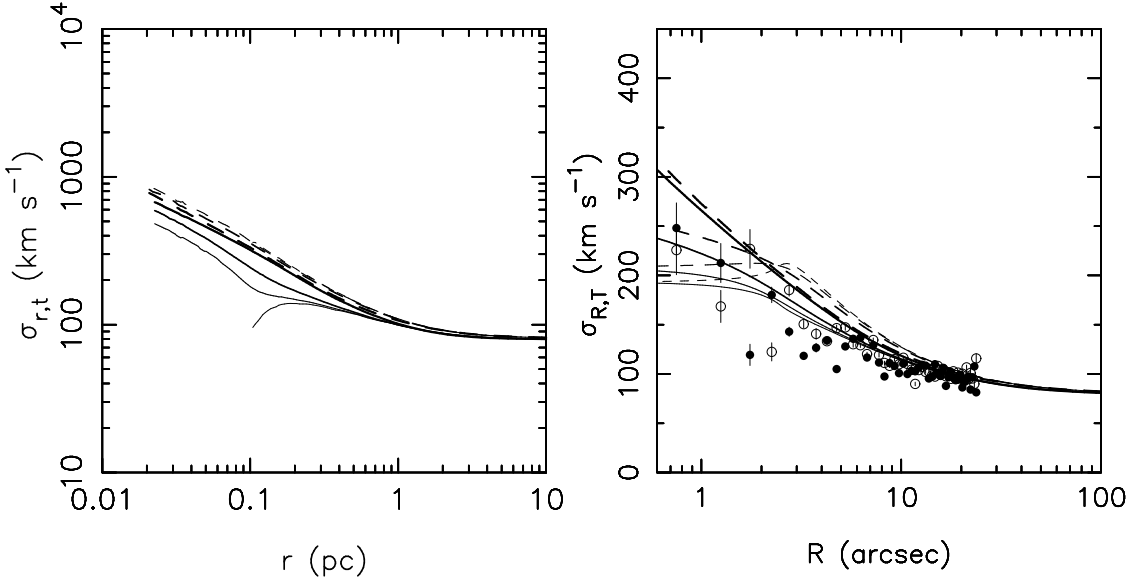}
\caption{Evolution of the space (left) and projected (right) velocity
dispersions for the same ``anisotropic core'' model illustrated in
Fig.~\ref{fig:rhoj}.
Solid (dashed) curves are radial (tangential) velocity dispersions.
Line thickness increases  with time; $t=(0,0.1,0.5,5)\times 10^9$ yr.
Symbols have same meaning as in Fig.~\ref{fig:anisot}.}
\label{fig:sigj}
\end{figure*}

In addition to being physically motivated,  
this approximation allows us to 
treat the evolution in $J$ and $E$ via
differential equations with just one space dimension, and
to formally separate the ``isotropization time'' from the
time scale for changes in the radial (i.e. energy) distribution.

The separation of time scales breaks down for certain orbits,
e.g. the lowest-energy orbits with $E\lap E_b$, and a full, 
$f(E,J,t)$
treatment could certainly be justified.
Here we note only that our initial conditions are somewhat
arbitrary, and that changes in the $E$-dependence of $f$ that 
would otherwise occur during the ``fast'' evolutionary phase could
be seen as establishing a slightly different $f(E)$ at the start of
 the ``slow'' evolutionary phase.

As the phase space density evolves, the configuration-space density 
$\rho(r)$ also changes, as well as the velocity dispersions 
$\sigma_r(r)$, $\sigma_t(r)$.
Our main constraint on these models is that the distribution of
stars, after $\sim 10$ Gyr, be consistent with the observed
distribution of late type stars at the Galactic center.
It is also reasonable to require that a model which matches the data now,
not be in such a rapid state of evolution that it
would quickly (in a time $\ll 10$ Gyr) evolve to a very
different form.

\subsection{Evolution in $J$}
\subsubsection{Equations}

The orbit-averaged Fokker-Planck equation describing changes in $f$
due to diffusion in $J$-space is
\cite[e.g.][]{Cohn:79}
\beq
{\partial N\over\partial t} = {\partial\over\partial {\cal R}}
\left( D_{\cal R}f + D_{\cal RR}{\partial f\over\partial {\cal R}}\right).
\label{eq:dNdt}
\eeq
Here ${\cal R}=J^2/J_c^2$ is a scaled angular momentum variable,
$J_c(E)$ is the angular momentum of a circular orbit of energy $E$,
$N(E,{\cal R}) = 4\pi^2P(E,{\cal R})J_c^2(E) f(E,{\cal R})$ is the
number density of stars in $(E,{\cal R})$ space, $P(E,{\cal R})$ is the
radial period of an orbit, 
and $\left\{D_{\cal R}, D_{\cal RR}\right\}$ are the angular momentum 
diffusion coefficients:
\begin{subequations}
\begin{eqnarray}
D_{\cal R}(E,{\cal R}) &=& -16\pi^2 {\cal R} r_c^2(E) \int {dr\over v_r} \left(1-{v_c^2\over v^2}\right) F_1(E,r), \\ 
D_{\cal RR}(E,{\cal R}) &=& {16\pi^2\over 3}{\cal R} 
\int {dr\over v_r} \bigg\{2{r^2\over v^2}\left[v_t^2\left({v^2\over v_c^2}-1\right)^2 + v_r^2\right]F_0(E) \nonumber \\
&+& 3{r^2v_r^2\over v^2} F_1(E,r) \nonumber \\
&+& {r^2\over v^2} \left[2v_t^2\left({v^2\over v_c^2}-1\right)^2 - v_r^2\right]F_2(E,r) \bigg\},
\end{eqnarray}
\label{eq:DR}
\end{subequations}
with
\begin{subequations}
\begin{eqnarray}
F_0(E) &=& 4\pi\Gamma\int_E^{\infty} dE'\overline{f}(E'), \\
F_1(E,r) &=& 4\pi\Gamma\int_{\phi(r)}^E dE'\overline{f}(E')
\left({E'-\phi\over E-\phi}\right)^{1/2}, \\
F_2(E,r) &=& 4\pi\Gamma\int_{\phi(r)}^E dE'\overline{f}(E')
\left({E'-\phi\over E-\phi}\right)^{3/2}
\end{eqnarray}
\label{eq:FF}
\end{subequations}
and $\Gamma\equiv 4\pi G^2m^2\ln\Lambda$.
In the expressions (\ref{eq:DR}), the integration interval 
is the radial range from periapsis to apoapsis.
Definitions for subsidiary variables can be found in Cohn (1979)
whose notation is adopted here.
Following Shapiro \& Marchant (1978) and 
Cohn \& Kulsrud (1978), the angular-momentum-averaged phase-space
density $\overline{f}$ that appears in equations~(\ref{eq:FF}) is
defined as
\beq
\overline{f}(E) = \int_0^1 d{\cal R} f(E,{\cal R}).
\label{eq:overline}
\eeq
Because we are ignoring changes in $E$, the function
$\overline f(E)$ does not change with time, nor
do the diffusion coefficients $\left\{D_{\cal R}, D_{\cal RR}\right\}$.

The practice of some authors
\cite[e.g.][]{MM:03} of approximating the ${\cal R}$-diffusion coefficients 
by their limiting values as ${\cal R}\rightarrow 0$ is not followed here.

\begin{figure}
\includegraphics[clip,width=0.425\textwidth]{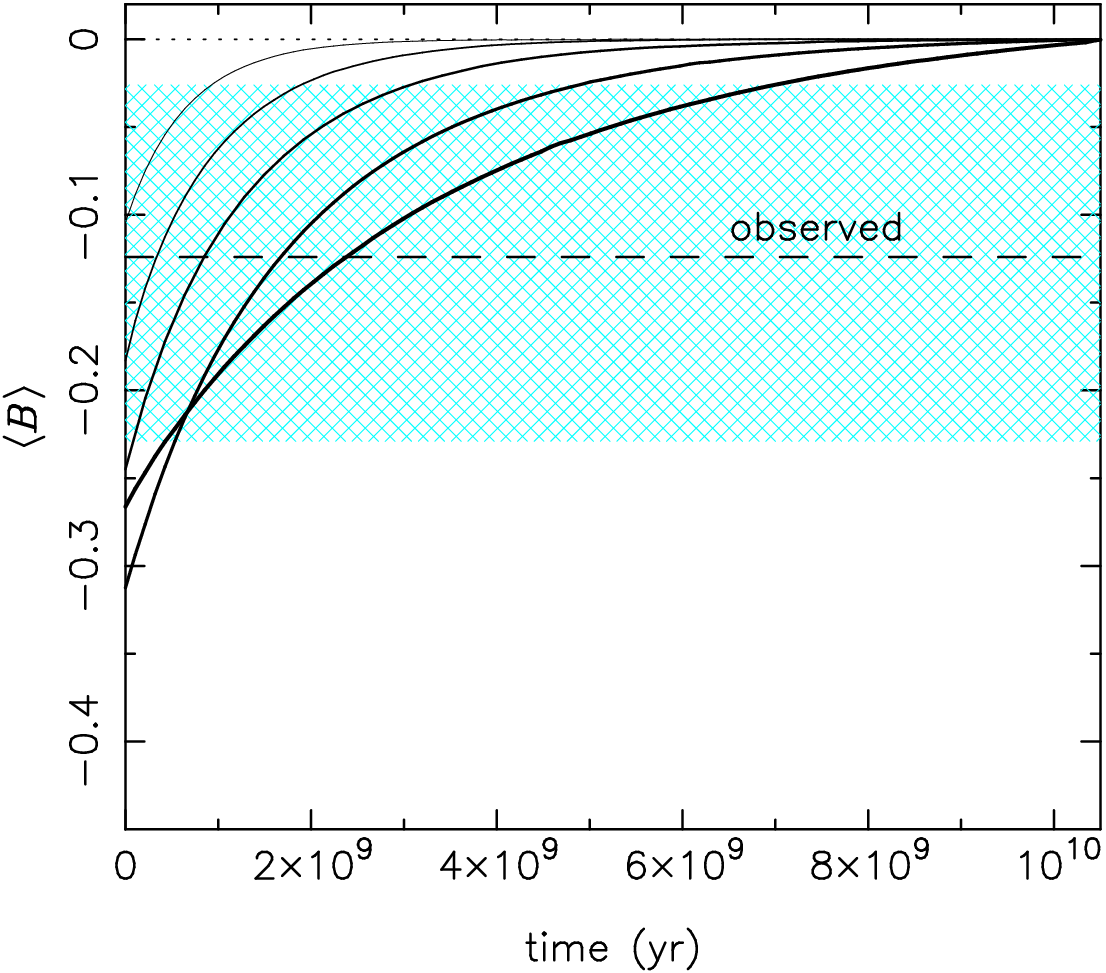}
\caption{Evolution of the projected anisotropy parameter $\langle{\cal B}\rangle$
defined in the text, computed over the projected annulus
$1''\le R\le 10''$.
Curves are from integrations starting from the five initial models illustrated in 
Figs.~\ref{fig:core2} and~\ref{fig:anisot}; increasing line
thickness denotes increasing values of $r_b$, the initial truncation
radius, from $0.02$ pc to $0.5$ pc.
Dashed line is the anisotropy at the Galactic center as computed
from the proper motion data of Sch\"odel et al. (2009), with
the 90\% confidence interval shown as the hatched (blue) region.
}
\label{fig:betas}
\end{figure}

Equation~(\ref{eq:dNdt}) was advanced in time numerically using the NAG
routine {\tt d03pcf}.
Initial conditions were $f(E,J)$
corresponding to the
anisotropic core models described in the previous section,
with various values of $r_b$ (e.g. Figs.~\ref{fig:core2}, ~\ref{fig:anisot}).

Scaling of the Fokker-Planck models to physical units of length and 
mass is fixed by the power-law density model, equation~(\ref{eq:rhoi}),
used to generate the initial conditions.
Since the Fokker-Planck equations are orbit-averaged, the relevant time unit is
the relaxation time.
In what follows, times will be expressed in years, 
based on a scaling that assumes a relaxation time given by
equation~(\ref{eq:tr_spitz2}) with $\tilde m=1\msun$ and $\ln\Lambda=15$.
If $\tilde m$ and $\ln\Lambda$ have different values than these.
the times given below should be multiplied by
\beq
\left({\tilde m\over 1\msun}\right)^{-1} 
\left({\ln\Lambda\over 15}\right)^{-1}.
\eeq

\subsubsection{Results}

Figure~\ref{fig:rhoj} shows the evolution of the space and projected
densities for the anisotropic core model with $r_b=0.1$ pc
(blue curves in Figs.~\ref{fig:core2},~\ref{fig:anisot}).
The central configuration-space ``hole'' is rapidly filled as 
the low-angular-momentum orbits are repopulated;
by a time of $\sim 10^9$ yr, the central density is essentially
unchanging inside $\sim r_b$.
Because (by assumption) there is no diffusion in $E$,
the central hole in {\it phase-} space remains, and so 
$\rho(r)$ evolves asymptoticall to the $\sim r^{-1/2}$ form demanded by
an isotropic $f(E)$ with a low-energy truncation
(Figure~\ref{fig:core1}).
The surface density evolves much less, since even at the
(projected) center, the surface density is dominated by
stars in the power-law envelope.
Thus, this model is a reasonable fit to the observed number
counts either at $t=0$ or at later times: the refilling of the
low-$J$ orbits does not greatly affect the observed densities.

Figure~\ref{fig:sigj} illustrates the evolution toward isotropy
in the same time integration.
Initially there is a strong velocity anisotropy near the center
due to the lack of eccentric orbits (Figure~\ref{fig:anisot}).
However the refilling of the low-$J$ orbits increases
$\sigma_r$ at $r\lap r_b$ on a time scale of $\sim 1$ Gyr
and the core is essentially isotropic thereafter.

Proper-motion data from  the (projected) inner parsec
of the Milky Way indicate a slight
degree of anisotropy \citep{Schoedel:09}.
We define an averaged anisotropy parameter $\langle{\cal B}\rangle$ 
as
\beq
\langle{\cal B}\rangle \equiv 
1 - {\langle\sigma_T^2\rangle\over\langle\sigma_R^2\rangle}
\eeq
where $\sigma_R$ and $\sigma_T$ are the radial and tangential
velocity dispersions  in the plane of the sky and
the $\langle\rangle$ denote number-weighted averages
over some radial range.
Adopting $1''\le R\le 10''$ for this range, the late type stars near
the Galactic center have
\beq
\langle {\cal B}\rangle = -0.124^{0.098}_{-1.05} 
\eeq
where the (90\%) confidence intervals were derived via the bootstrap.
Figure~\ref{fig:betas} compares the observed value of $\langle{\cal B}\rangle$
with the values predicted by the evolving models.
The 90\% observed upper bound, $\langle{\cal B}\rangle\approx -0.026$, 
is almost consistent with isotropy, which is the asymptotic state of
the time-dependent models; 
while the observed lower bound, $\langle{\cal B}\rangle\approx -0.23$,
is almost as low as the initial anisotropy of the most
extreme core model considered here.
Thus, the Fokker-Planck models remain consistent with the observed 
degree of anisotropy over essentially the entire time interval and for
a wide range of initial conditions.

Figure~\ref{fig:betas} also gives an indication of how the 
time to establish isotropy in an  initially anisotropic
core varies with the size of the core.
Larger values of $r_b$ imply both a higher initial anisotropy,
and a longer time scale for the establishment of isotropy.

\subsubsection{Summary}

Cores formed by the exclusion of  small-periapse  orbits
evolve toward isotropy on a $\sim$ Gyr timescale.
This evolution does not produce great changes in the observable
properties of the core, either in  the density or the velocity
dispersions.
The anisotropy observed at the Galactic center
is consistent with the evolving models at both early and late times.

\subsection{Evolution in $E$}

Evolution in $E$-space drives $f$ toward the
quasi-steady-state form
\beq
f\sim |E|^{1/4},\ \ \ \ \rho \sim r^{-7/4}
\label{eq:BW}
\eeq
at $E\lap -G\mh/\rh$ and $r\lap\rh$, 
on a time scale that is roughly the relaxation time 
at $\rh$ \citep{BW:76,LS:77}.
Equation~(\ref{eq:BW}) corresponds to a zero net flux in $E$-space near
the hole.
In reality, loss of stars into the SMBH implies a non-zero flux,
causing a gradual evolution (expansion) of the cluster, although
without much change in the form of $\rho(r)$ \cite[e.g.][]{SM:78,MCD:91}.
We ignore that complication here since the density in our models
near the SMBH remains far below that of the quasi-steady-state models
at most times, implying a very small flux into the SMBH.

Many authors have explored quasi-steady-state solutions
to this equation and to the more general equations that allow
for a dependence of $f$ on orbital angular momentum and stellar
mass \cite[as reviewed by][]{Merritt:06b}.
After a {\it finite} time $\lap t_r$, the form of $f(E)$ will still reflect
the initial conditions.
Quinlan (1996) emphasized this in the case of stellar systems
without central SMBHs.
Freitag et al. (2006) explored in a limited way how the structure
of Galactic center models depends on the assumed initial density
profile.
They considered initial profiles $\rho\sim r^{-\gamma}$ with $\gamma$
as small as $0.75$ (steeper, i.e. closer to the asymptotic Bahcall-Wolf form,
than the steepest initial profiles considered here).
Freitag et al. found that a time of order $t_r(\rh)$ is required to
erase details of the initial conditions.
Based on Figure~\ref{fig:tr} that time is $\sim 20$ Gyr.

\subsubsection{Equations}

The orbit-averaged, isotropic Fokker-Planck equation is
\begin{subequations}
\begin{eqnarray}
& &{\partial N\over\partial t} = {\partial\over\partial E}
\left(D_{EE}{\partial f\over\partial E} + D_Ef\right),\\
& & D_{EE}(E) = 16\pi^3\Gamma\times  \\ 
& & \left[q(E) \int_{-\infty}^E dE'f(E') + 
\int_E^\infty dE' q(E')f(E')\right], \nonumber \\
& & D_E(E) = 16\pi^3\Gamma\int_E^\infty dE'p(E')f(E')
\end{eqnarray}
\label{eq:fp}
\end{subequations}
\noindent
(e.g. Cohn 1980; Spitzer 1987).
Here $N(E)=4\pi^2p(E)$ is the number density of stars in energy
space and $p(E) = 4\sqrt{2}\int r^2 \sqrt{E-\phi(r)} dr
=\partial q/\partial E$ is a phases-space volume element.
The functions $f$, $D_E$ and $D_{EE}$ are understood to
depend on time; as above, the gravitional potential is assumed
to remain fixed, as do the functions $p$ and $q$.

Equations~(\ref{eq:fp}) were solved using the NAG routine {\tt d03pcf}.
The $E$-space flux was set to zero at the inner 
boundary of the energy grid, as justified above.
Two sorts of initial conditions were considered: 
(1) $f(E)$ corresponding to an ``isotropic core'' model;
(2) ${\cal R}$-averaged $f$'s from the final time steps of the $J$-integrations described
in the previous section, which started from ``anisotropic core''
models.
(The final $f$'s from these integrations were almost precisely
isotropic; cf. Figure~\ref{fig:betas}.)
In both cases, the initial conditions will be labelled in terms
of $r_b$.

\begin{figure}
\includegraphics[clip,width=0.425\textwidth]{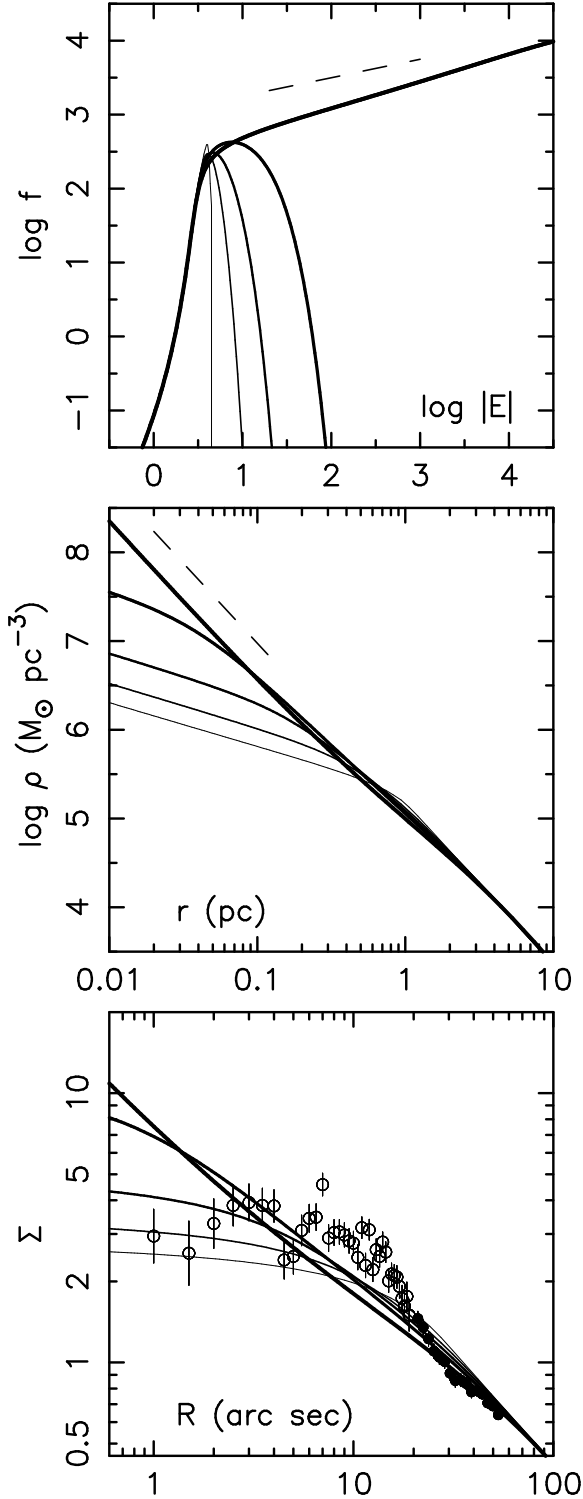}
\caption{Evolution of the phase-space density (top),
configuration-space density (middle), and surface density
(bottom) of the ``isotropic
core'' model with $r_b=1$ pc (Fig.~\ref{fig:core1}).
Increasing line thickness denotes increasing time,
$t=(0,0.2,0.5,1,2)\times 10^{10}$ yr.
Dashed lines are the asymptotic forms
for $f$ and $\rho$, i.e. $f\sim |E|^{1/4}$,
$\rho\sim r^{-7/4}$.
$E$, $f$ and $\Sigma$ are in arbitrary units.
}
\label{fig:frhosig}
\end{figure}

\subsubsection{Results}

Figure~\ref{fig:frhosig} shows the evolution  of
the ``isotropic core'' model with $r_b=1$ pc.
Diffusion in $E$-space causes stars to gradually occupy
orbits of lower (more bound) energies.
However even after $10$ Gyr -- roughly the relaxation
time at $r_b$ (Figure~\ref{fig:tr}) -- 
$f$ and $\rho$ are still far from their steady-state forms
at low energies/small radii.
The Bahcall-Wolf solution is only reached
after a time that is roughly twice as long.

The lower panel of Figure~\ref{fig:frhosig} highlights an 
interesting coincidence.
All of the models considered here have (by assumption) a density that obeys
\beq
\rho \sim r^{-1.8}
\eeq
at large radii -- the observed dependence of the density of old stars
on radius beyond $\sim 1$ pc.
But this is essentially the same slope as in the steady-state
Bahcall-Wolf profile, $\rho\sim r^{-1.75}$ which is the asymptotic
form of $\rho(r)$ at {\it small} radii.
As long as the initial core radius is smaller than $\sim \rh$,
it follows that $\rho(r)$ will
evolve in an approximately self-similar way: the core will shrink,
while outside the core, $\rho(r)$ will continue to obey $\rho\sim r^{-1.8}$.
Reproducing the observed density profile is therefore 
simply a matter of choosing the appropriate, initial value of $r_b$.

\begin{figure}
\includegraphics[clip,width=0.425\textwidth]{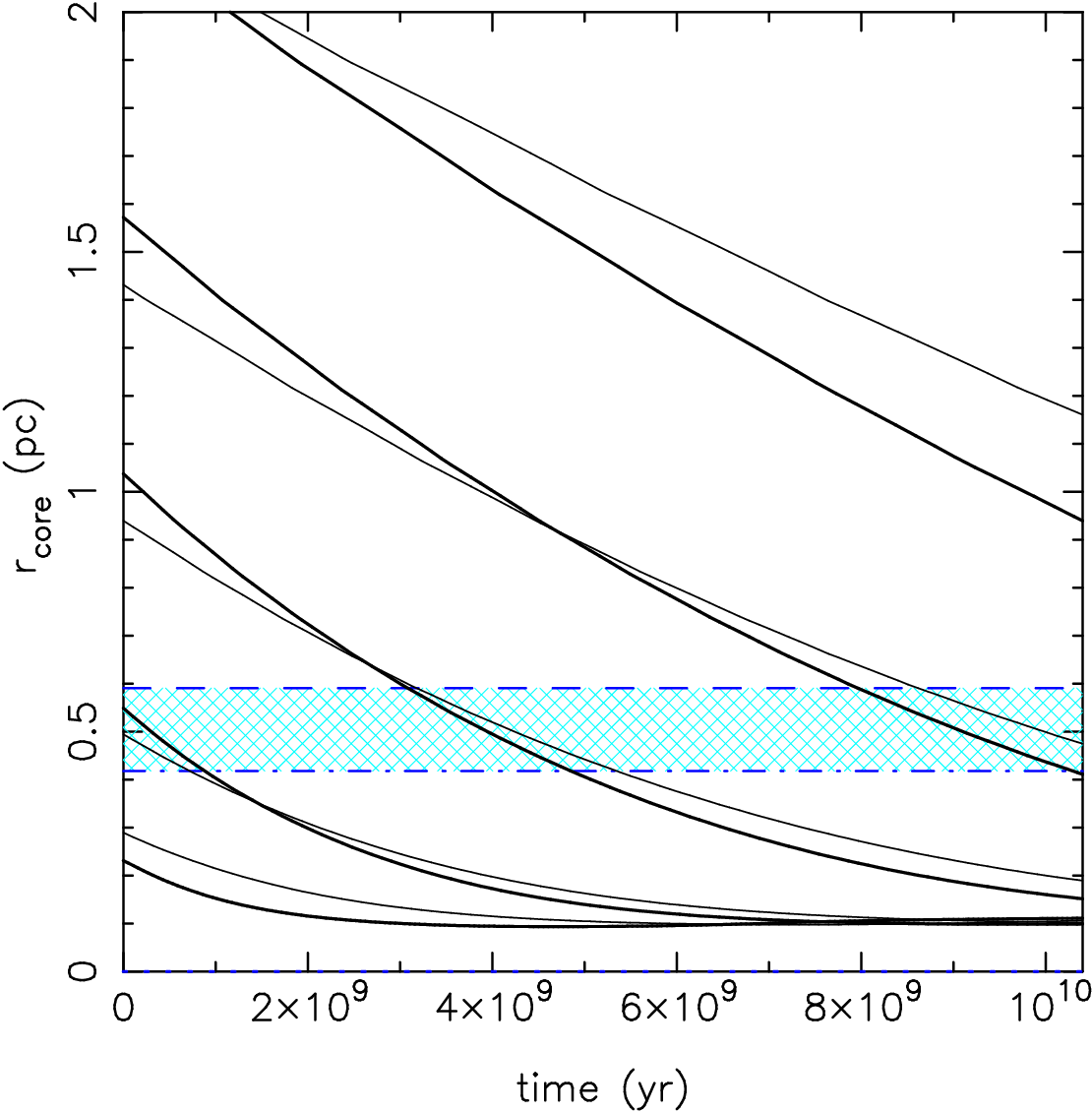}
\caption{Evolution of the core radius in various models.
{\it Thick curves:} ``isotropic-core'' models.
{\it Thin curves:} ``anisotropic-core'' models.
The hatched (blue)  region encloses the two estimates
of the Milky Way core radius, as discussed in the text.
}
\label{fig:core}
\end{figure}

Which values of $r_b$ give cores of the right size now?
Two estimates were derived in \S2 for the core radius of the old 
stellar population, $0.59$ and $0.42$ pc.
These values are plotted in Figure~\ref{fig:core},
which also shows core radii as a function of time in the evolving
models (computed in the same way, i.e. by finding the projected
radius at which the surface density falls to 1/2 of its value at
$0.04$ pc).
The figure suggests that the currently observed
core is consistent with initial cores having sizes in the range
\beq
1 {\rm pc} \lap \rc \lap 1.5 {\rm pc},
\eeq
or $2-3$ times the current value.
These values are comparable with the SMBH influence radius
(Fig.~\ref{fig:tr}).

\section{Segregation of the massive remnants}

Old stellar populations contain remnants: 
white dwarves (WDs), neutron stars (NSs), 
and stellar-mass black holes (BHs),
the end-products of stars with initial masses
$1-8\msun$ (WDs), $ 8-30\msun$ (NSs), and
$\sim 30-100\msun$ (BHs). 
Standard assumptions about the initial mass function imply 
that $\sim 1\%$ of the total mass of an old 
population should be in the form of stellar BHs \citep{Alexander:05}, 
although observational constraints on the BH number density
near the Galactic center are weak \cite[e.g.][]{Muno:05}.

The stellar BHs have significantly higher masses ($\sim\times 10$)
than either the main-sequence stars or the other types of remnant
that collectively dominate the total mass density. 
The BHs should therefore lose orbital energy due to dynamical friction
and congregate around the SMBH.
Assuming that the total density obeys an expression similar
to equation~(\ref{eq:rhoi}), i.e. $\rho\sim r^{-2}$,
the time for a $10\msun$
BH on a circular orbit to spiral all the way in to the center is
less than 10 Gyr for a starting radius inside $4-5$ pc
\citep{Morris:93,MG:00}.
Depending on the assumed mass fraction in BHs, and on their
initial density profile, the mass density of BHs 
after $5-10$ Gyr is predicted to
match or exceed that of the other populations inside $\sim 10^{-2}$ pc.
If this occurs, the BHs will undergo gravitational scattering
from themselves and from the other populations,
leading to a quasi-steady-state 
$n\sim r^{-2}$ density profile in the innermost regions,
and to a slightly shallower profile in the lighter mass components
 \citep{Freitag:06a,HA:06a,AH:09}.

All of the studies cited above assumed or derived a total mass density
that increases as $\rho\sim r^{-\gamma}$, $1.3\lap\gamma\lap 2.3$
down to $\sim 10^{-5}$ pc from the SMBH.
If instead the dominant population has a core,
the dynamical friction force will increase more slowly toward
the center inside $\sim \rc$, implying somewhat longer inspiral times.

But there is potentially an even more dramatic way in which a core
can affect the rate of orbital decay, as we now show.
The instantaneous frictional force felt by a (massive)
test body of mass $\mbh$ and velocity $\mathbf{v}$ is
\beq
\mathbf{a} = -\frac{4\pi G^2 \mbh\rho(r)F(v)\ln\Lambda} {v^3} \mathbf{v}
\label{eq:adf}
\eeq
where $F(v)$ is the fraction of stars locally
that are moving more slowly than $v$
.
If the latter are described by an isotropic $f$, then
\beq
\rho(r)F(v) = 4\sqrt{2}\pi \int_{\phi(r)}^{v^2/2+\phi(r)}
dE f(E) \sqrt{E-\phi(r)}.
\eeq
If in addition $f$ is truncated at energies below $E_b$ -- 
an ``isotropic core'' -- then $F$ falls to zero for orbits with
energies
\beq
{v^2\over 2} + \phi(r) \le E_b 
\eeq
since there are {\it no} stars locally that move slower than $v$ 
for these energies --
even if the configuration-space density is nonzero.
Assuming a circular orbit for the test body, and that the orbit lies inside
the influence radius of the SMBH, this condition becomes
\beq
r \lap {r_b\over 2}
\eeq
with $E_b=\phi(r_b)$.
Thus, inside  $\sim 1/2$ the core radius, the frictional force
drops precisely to zero.

This result is not simply an artifact of the brute-force truncation 
of $f$.
Consider a core in which $\rho\propto r^{-\gamma}$.
In a point-mass potential, this density is reproduced by
\beq
f(E) = f_0 \left|E\right|^{\gamma-3/2}
\label{eq:fpl}
\eeq
and $F$ for a circular orbit is easily shown to be
\beq
F(\gamma) = {2\over\sqrt{\pi}} {\Gamma(\gamma+1)\over\Gamma(\gamma-1/2)}
\int_{1/2}^1 dz z^{\gamma-3/2}\sqrt{1-z}.
\eeq
This function varies smoothly from $F=0.5$ at $\gamma=2$,
the singular isothermal sphere, to $F=0$ at $\gamma=0.5$.

\begin{figure}
\includegraphics[clip,width=0.425\textwidth]{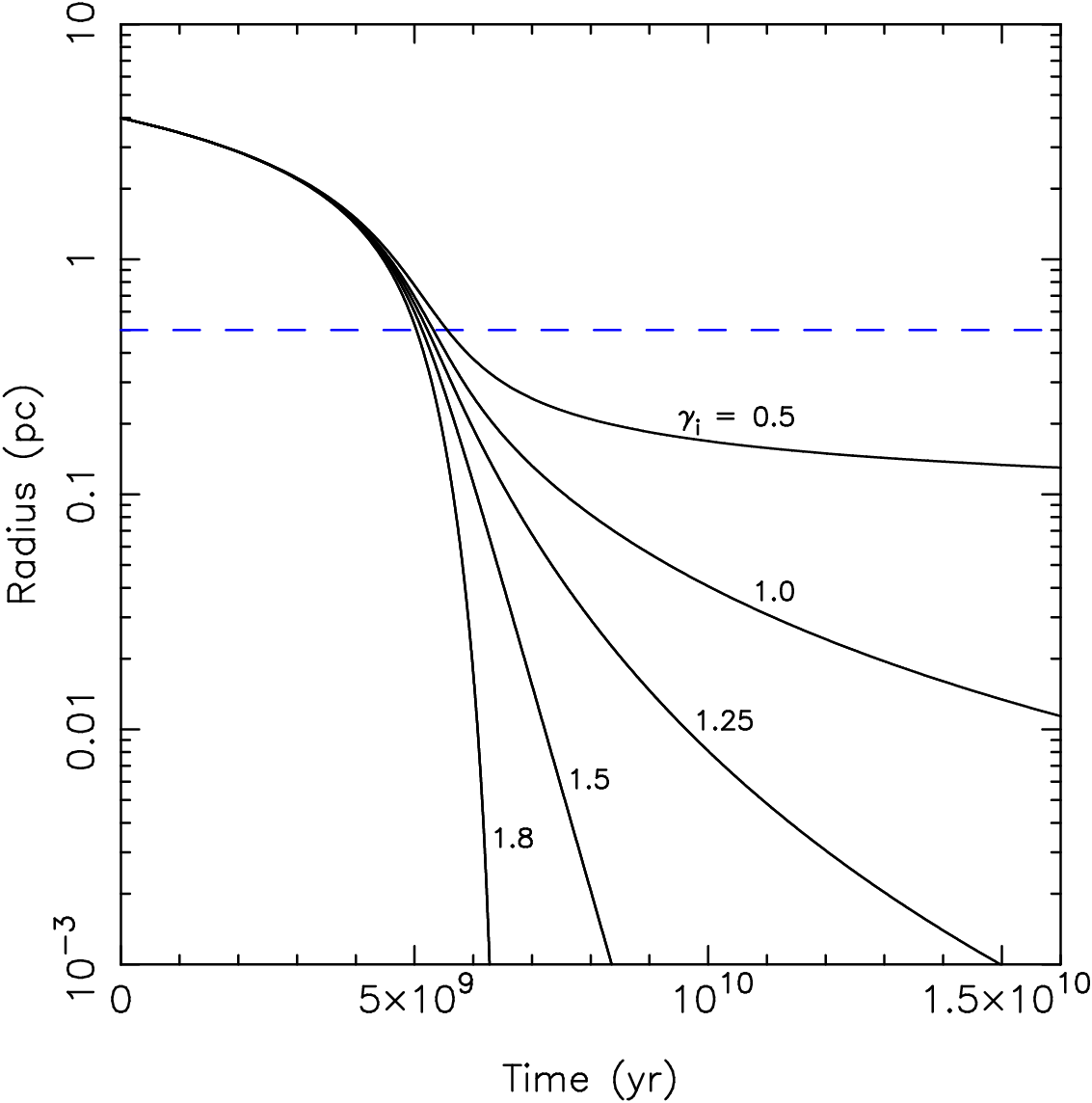}
\caption{Trajectories of $10 \msun$ BHs as they spiral in to the
Galactic center on circular orbits, starting from a radius of
$4$ pc.
The assumed background density is a power-law, $\rho\propto r^{-1.8}$
at large radii, with an inner core of radius $r_0$, as described
by equation~(\ref{eq:nofr}) with $\alpha=2$; $r_0$ was set to $0.5$ pc
(dashed/blue line) and the inner power-law slope of the density 
was varied, as indicated.
}
\label{fig:inspiral}
\end{figure}

As an even more general illustration of this effect,
we computed the evolution of a circular orbit 
in a background density described
by the broken power-law model of equation~(\ref{eq:nofr})
with $\alpha=2$.
The outer slope was fixed at $\gamma=1.8$ and 
the normalization was chosen to reproduce the density of the 
``fiducial'' model, equation~(\ref{eq:rhoi}), outside the core.
Setting the inner slope $\gamma_i$ to $1.8$ gives a 
model similar to those assumed in the studies cited above; 
while small values of $\gamma_i$ correspond to a core.
The radius $r$ of a test body's orbit decays as
\beq
{1\over r}{dr\over dt} = 
-\left|\mathbf{a}\right| \left({dJ\over dr}\right)^{-1}.
\eeq
The function $F(v)$ was computed using the expression \citep{Szell:05}
\begin{eqnarray}
&&F(v) = 1 - {1\over \rho}\int_{E}^0 d\phi' {d\rho\over d\phi'} 
\times \nonumber \\
&&\left\{ 1 + {2\over\pi} \left[{v/\sqrt{2}\over\sqrt{\phi'-E}} - 
\tan^{-1}\left({v/\sqrt{2}\over\sqrt{\phi'-E}}\right)\right]\right\}.
\end{eqnarray}
Figure~\ref{fig:inspiral} shows the trajectories of $10\msun$ BHs
starting from a distance of $4$ pc, assuming various values for the
inner density slope; in each case we set the core radius parameter
$r_0$ to $0.5$ pc.
As predicted, the rate of orbital decay begins to slow when 
$r\lap r_0$ for small values of $\gamma_i$, and the decay essentially
stalls, at $r\lap r_0/2$ pc, when $\gamma_i=0.5$.
We stress that the configuration-space density is nonzero at all radii
in these models, and in fact increases monotonically toward the center;
the precipitous drop in the frictional force is
due to the lack of low-velocity stars in the core when $\gamma_i$ is small.

Another consequence of a core is that the dynamical
friction force along an orbit is not so strongly peaked near periapse,
which is a necessary condition for an orbit to circularize.
In the presence of a core, the distribution of BH orbits will
therefore remain more nearly isotropic (assuming that it starts out
isotropic).
This, combined with the cutoff in the dynamical friction
force at $r\approx r_0$, suggests that the evolved BH density would 
rise rapidly toward $r\approx r_0$, then
follow $\sim r^{-0.5}$ toward smaller radii, the density law of an istropic
population with an inner hole in phase space.
If the background density is itself evolving,
as in the models of the previous section, the core radius of the
population that produces the dynamical friction force
will decrease with time, causing the radius of peak BH density to
also migrate inward on the same time scale.

\subsection{Equations}

As in previous sections, the evolution of the distribution of
stellar BHs was followed using the orbit-averaged Fokker-Planck
equation.
Define $f_{BH}(E,{\cal R})$ to be the number density in phase space 
of a population of massive objects (BHs), of individual mass $\mbh$.
We assume that the associated mass density $\rho_{BH}$ is small compared with
the total mass density $\rho$ due to stars and (less massive) remnants.
Alexander \& Hopman (2009) show that the limiting density ratio
for ignoring the self-interaction of the BHs is
$\rho_\mathrm{BH}/\rho_\star<m_\star/m_\mathrm{BH}\approx 0.1$.
This condition is violated at late times in some of the integrations
described below; this effectively defines the maximum time 
at which the solutions are valid.

The orbit-averaged equation describing the evolution of $f_{BH}$
as the massive objects experience dynamical friction
against the background of less massive objects is
\beq
{\partial N_M\over\partial t} = 
{\partial\over\partial E}\left({D_E} f_{BH} \right) +
{\partial\over\partial {\cal R}}\left( {D_{\cal R}} f_{BH} \right)
\label{eq:df}
\eeq
where $N_M(E,{\cal R}) = 4\pi^2P(E,{\cal R})J_c^2(E) f_{BH}(E,{\cal R})$ is the
number density in $(E,{\cal R})$ space as before.
The diffusion coefficients depend on the (possibly time-dependent) 
distribution of 
low-mass objects.
Let $f_i(E,{\cal R},t)$ be the phase-space number density of stars
with mass $m_i$, $m_i\ll \mbh$. 
Then  \cite[e.g.][]{Takahashi:97}
\begin{subequations}
\begin{eqnarray}
D_E(E,{\cal R}) &=& -8\pi^2 \mbh J_c^2(E) 
\sum_i m_i 
\int {dr\over v_r} F_{1i}(E,r), \\ 
D_{\cal R}(E,{\cal R}) &=& -16\pi^2 \mbh{\cal R} r_c^2(E) 
\sum_i m_i 
\int {dr\over v_r} \left(1-{v_c^2\over v^2}\right) F_{1i}(E,r), 
\end{eqnarray}
\label{eq:xx}
\end{subequations}
with
\begin{equation}
F_{1i}(E,r) = 4\pi\gamma\int_{\phi(r)}^E dE'\overline{f}_i(E',t)
\left({E'-\phi\over E-\phi}\right)^{1/2}
\label{eq:yy}
\end{equation}
and $\gamma\equiv 4\pi G^2\ln\Lambda$; $\overline{f}_i$
is the angular-momentum-averaged $f$ as defined above.

Defining the phase-space mass densities of BHs and the other
populations (collectively referred to, henceforth, as ``the stars'')
respectively as
\begin{equation}
g_{BH} = \mbh f_{BH}, \ \ \ \ g = \sum_i m_if_i,
\end{equation}
the evolution equation for the BHs can be written
\begin{eqnarray}
&&{PJ_c^2\over 8\pi\gamma \mbh}{\partial g_{BH}\over\partial t} = \nonumber \\
&&-{\partial\over\partial E}
\left[g_{BH} J_c^2 \int {dr\over v_r} \int_{\phi(r)}^E dE'\overline{g}(E',t)
\left({E'-\phi\over E-\phi}\right)^{1/2}\right] \nonumber \\
&&-2{\partial\over\partial {\cal R}}
\left[g_{BH} {\cal R}r_c^2 \int {dr\over v_r} \left(1-{v_c^2\over v^2}\right)
\int_{\phi(r)}^E dE'\overline{g}(E',t)
\left({E'-\phi\over E-\phi}\right)^{1/2}\right].
\label{eq:fpbh}
\end{eqnarray}

This equation was integrated forward using an explicit scheme with
second-order space derivatives and first-order time derivative.
Two choices were considered for the stellar phase-space density $g$: 
(1) a time-independent model with an ``anistropic core'';
(2) a time-dependent, isotropic model in which $g$ evolves according
to equation~(\ref{eq:xx}).
In both cases, the phase-space density of the BHs was assumed to be
the same as that of the stars at $t=0$.

If the mass density in stellar BHs becomes comparable to that of the stars
at any radius,
equations (\ref{eq:fpbh}) are no longer valid, since the BHs will begin to feel
perturbations from each other and because the stellar distribution
will be affected by heating from the BHs.
Whether, or when, this occurs depends on the assumed initial
normalization of $\rho_{BH}$.
In plots that follow,
BHs were assumed to be a fraction $10^{-2}$ of the 
total mass density initially.

Results in this section are expressed in years, 
assuming that $\mbh=10\msun$.
Times can be scaled to different values of $\mbh$ using the 
simple proportionality of the dynamical friction force on $\mbh$
(eq.~\ref{eq:adf}).

\subsection{Results}

\begin{figure}
\includegraphics[clip,width=0.425\textwidth]{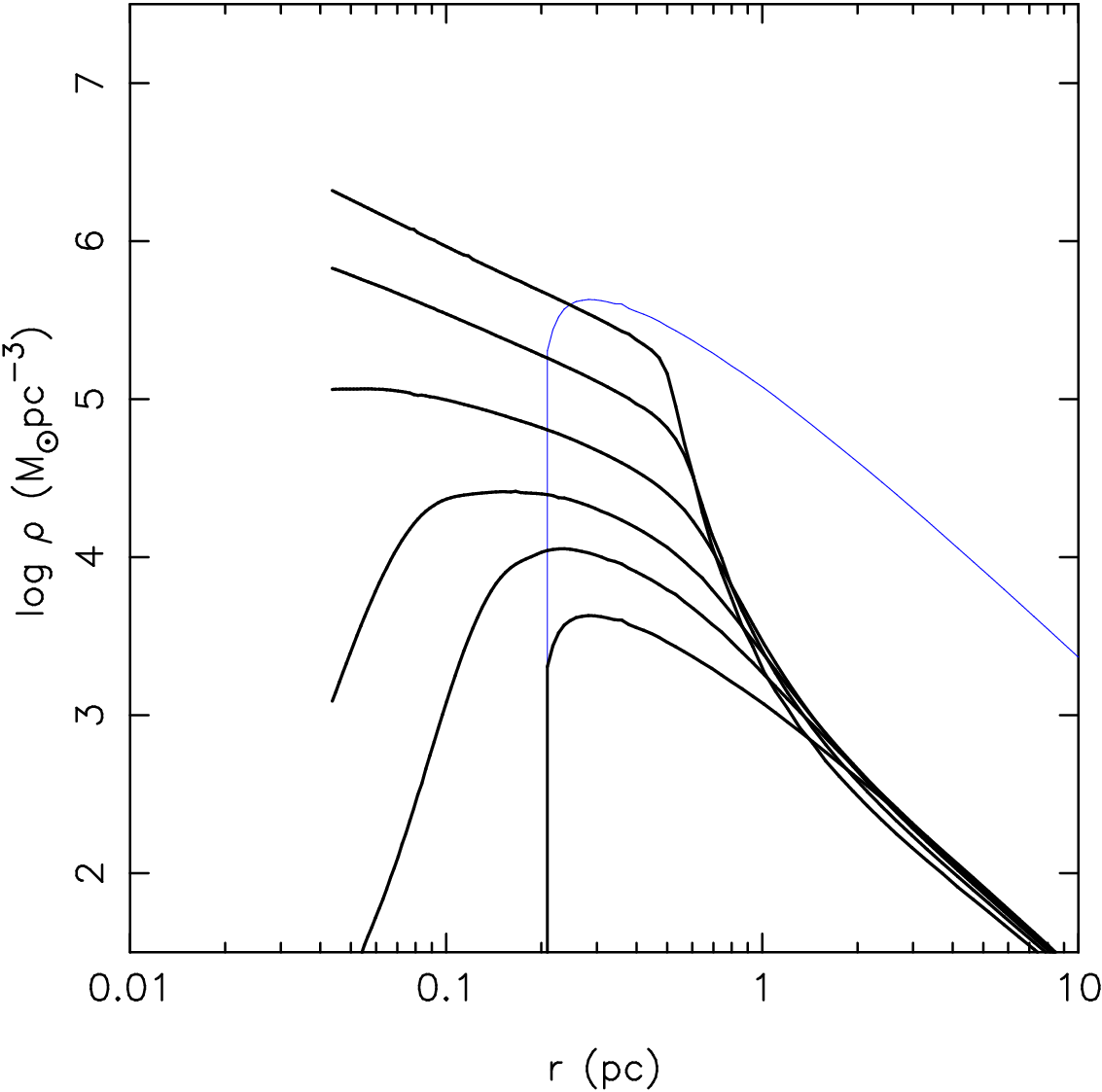}
\caption{Evolution of the density of stellar BHs assuming
a fixed background density, shown by the thin (blue) line.
Curves show the density at times ($0,1,2,4,8,16$) Gyr.}
\label{fig:seg1}
\end{figure}

Figure~\ref{fig:seg1} shows the evolution of the BH mass density
assuming a fixed stellar background,
 corresponding to an ``anisotropic core'' model with $r_b=0.2$ pc.
The initial BH distribution is likewise anisotropic.
As time progresses, the density of BHs drops at a radius of a few parsecs
and rises inside $\sim 1$ pc; BHs accumulate at energies
near the core due to the falloff in the dynamical friction force there,
as discussed above.
At late times ($\gap 5$ Gyr), the BH velocity distribution 
is slightly biased toward circular motions beyond $\sim 1$ pc,
and toward radial motions inside $\sim 1$ pc;
as noted above, dynamical friction in the
presence of a core does not efficiently circularize orbits.
To a reasonable approximation, the phase-space distribution of the BHs
at late times is isotropic,
and because the {\it stellar} distribution has not been allowed to evolve,
the BH density remains zero on orbits with energies below 
$E_c\approx E_b/2$.
The result is a $\rho_{BH}\sim r^{-0.5}$ cusp at $r\lap 2r_b\approx 0.4$ pc.
Note that the density of BHs just approaches that of the stars
at the final time ($16$ Gyr) in this integration.

\begin{figure}
\includegraphics[clip,width=0.425\textwidth]{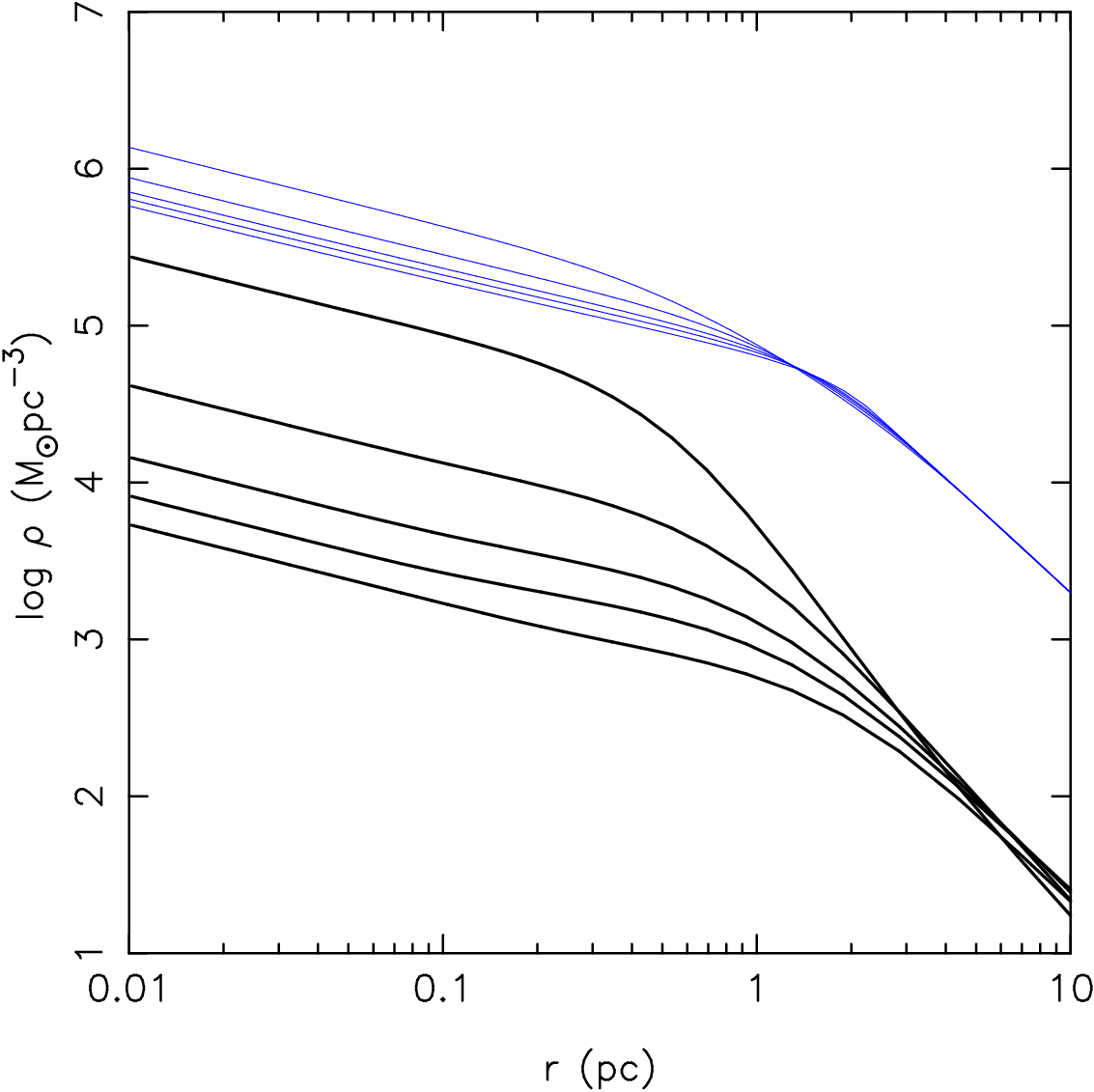}
\caption{Evolution of the number density of stellar BHs 
(thick/black curves) assuming
an evolving background (stellar) density (thin/blue curves),
starting from an ``isotropic core'' model with $r_b=2$ pc 
in both components.
Times shown are ($0,1,2,4,8$) Gyr.
}
\label{fig:seg2}
\end{figure}

\begin{figure}
\includegraphics[clip,width=0.425\textwidth]{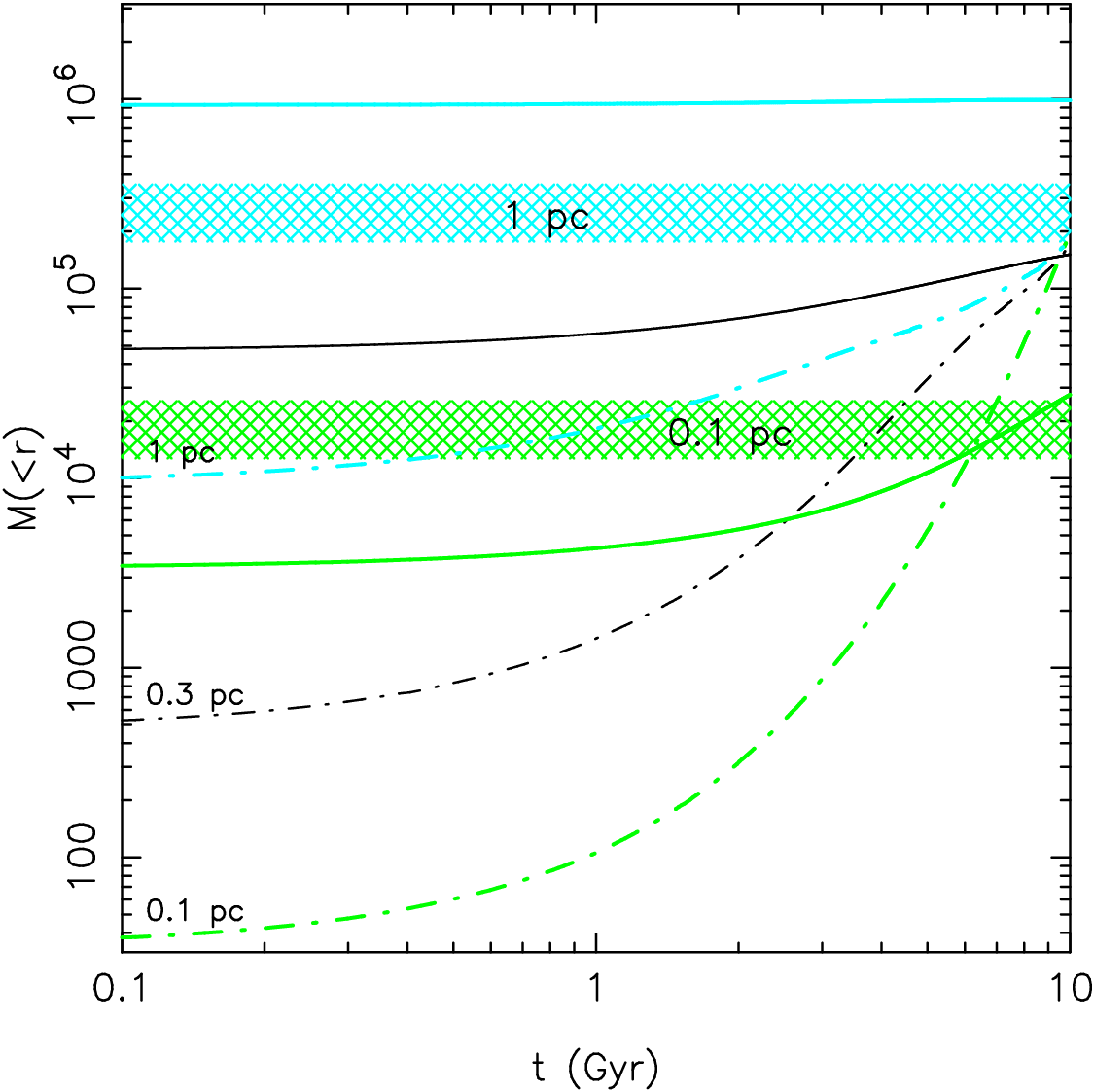}
\caption{Evolution of the enclosed mass in stellar BHs
(dash-dotted lines) and stars (solid lines) in an integration
like that of Figure~\ref{fig:seg2}, with $r_b=0.1$ pc.
Curves show mass enclosed within ($0.1, 0.3, 1.0$) pc
assuming that stellar BHs have initially 1\% the mass density
of stars at each radius.
The hatched regions show the estimates of Miralda-Escude \& Gould
(2000) for the mass of stellar BHs within 1 pc after 10 Gyr,
assuming dynamical friction against a fixed stellar density
cusp;
and of Hopman \& Alexander (2006a) for the mass of BHs within
0.1 pc based on steady-state Fokker-Planck solutions.
The vertical width of each hatched region corresponds to an
(arbitrary) factor two in mass.
}
\label{fig:mofr}
\end{figure}

Allowing the stellar distribution to evolve is more realistic.
Figures~\ref{fig:seg2}-\ref{fig:mofr} show the results of two 
such integrations.
In Figure~\ref{fig:seg2}, 
he initial distributions of stars and BHs were generated
from an ``isotropic core'' model with $r_b=2$ pc.
The stellar distribution was allowed to evolve according to 
equations~(\ref{eq:fp}),
yielding a time-dependent $g(E,t)$ which was inserted into
equations~(\ref{eq:fpbh}) at each time step to compute the diffusion
coefficients acting on $g_{BH}$.
In these integrations, the stellar core shrinks, on a time scale 
that is $\sim 10$ times longer
than the time scale for the BHs to accumulate around the core.
As a result, the BHs ``follow'' the stellar core inward.
Their density does not rise so steeply toward the core
as in the integrations with fixed stellar density
(Figure~\ref{fig:seg1}) since the radius at which they would
otherwise accumulate changes with time.
Figure~\ref{fig:mofr} shows the mass enclosed within ($0.1, 0.3, 1$)pc 
vs. time for both components, in a second integration starting
from $r_b=1$ pc.
Also shown for comparison is the mass in BHs estimated by Miralda-Escude
\& Gould (2000) to lie within the central parsec after $10$ Gyr, 
assuming dynamical friction against a fixed stellar background;
and the mass in BHs estimated by Hopman \& Alexander (2006) to
lie within 0.1 pc, based on their steady-state multi-mass Fokker-Planck
solutions.

The feature that we wish to emphasize here is the sensitivity,
in our models, of the final density in BHs to the elapsed time.
Figures~\ref{fig:seg2} and~\ref{fig:mofr} suggest that it would be 
unjustified to assume that the stellar BHs have reached a steady-state
density by now at any radius inside $\sim 1$ pc.
This is even more true if star formation has been an ongoing process 
in the nuclear star cluster \citep[e.g.][]{SM:96,Figer:04}, since
the mean age of stars and remnants may be much less than 10 Gyr
(\S\ref{sec:starformation}).

\begin{figure}
\includegraphics[clip,width=0.425\textwidth]{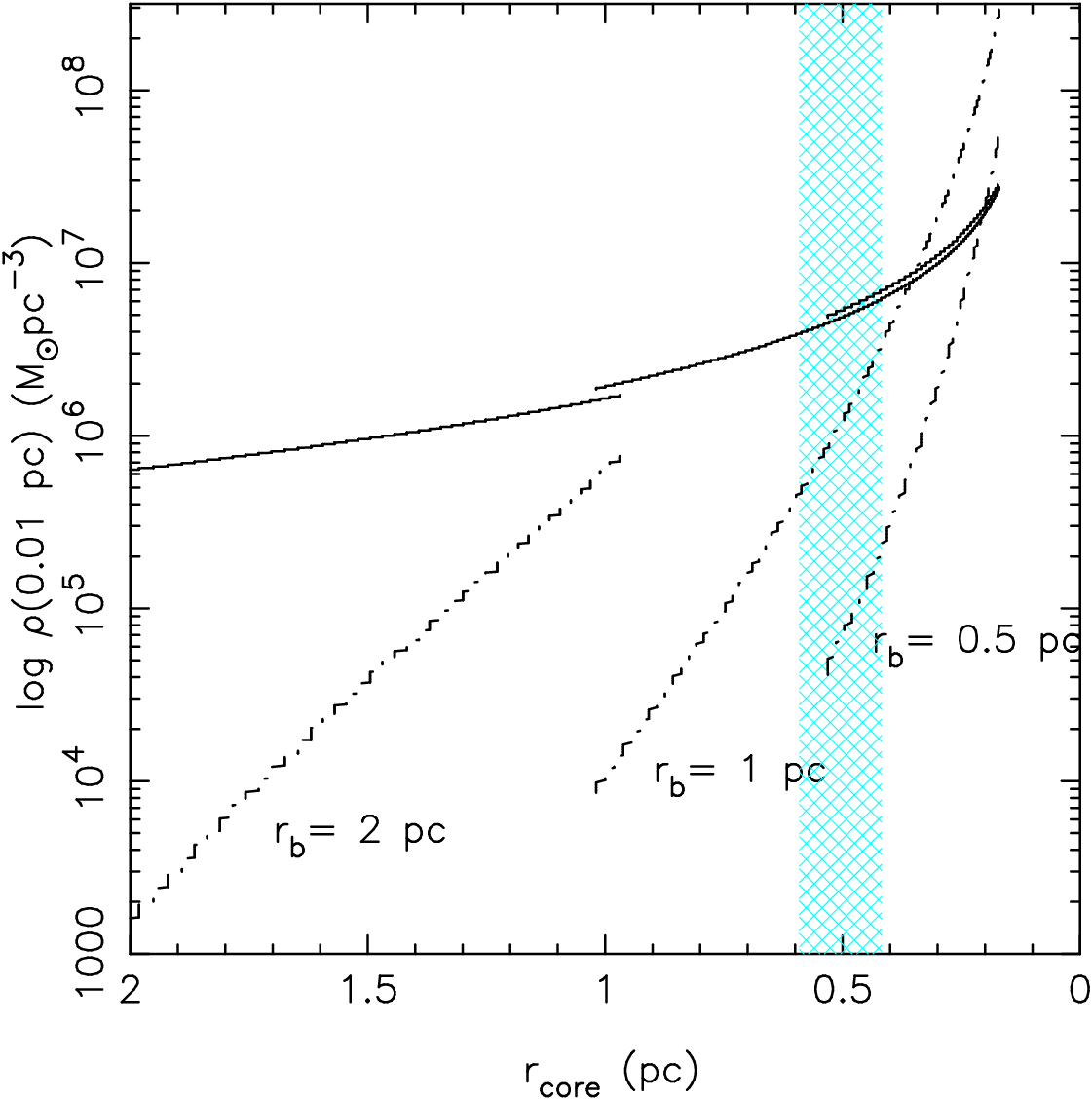}
\caption{Density of stellar BHs (dash-dotted curves) and stars 
(solid curves) at $r=0.01$ pc
in three models with cores of different initial size,
$r_b=(0.5,1,2)$ pc.
Horizontal axis is core radius of the stars, which decreases
with time as the core shrinks.
Integration times are 10 Gyr.
Vertical hatched region indicates the radius of the observed core.
}
\label{fig:rhorcore}
\end{figure}

This point is made more forcefully in 
Figure~\ref{fig:rhorcore}, which compares stellar and BH densities
at $r=0.01$ pc in three models with different initial core
sizes.
The figure shows that when the evolving stellar core
reaches a size consistent with the observed core,
the density in stellar BHs can be substantially less
than that of the stars at this radius.
In the steady-state models, the BH density meets or exceeds that
of the stars at radii of $10^{-2}-10^{-1}$ pc
\citep[e.g.][Fig.~10]{Freitag:06a},
and gravitational wave inspiral of BHs is dominated by BHs
at these distances \citep{HB:95,HA:05}.
The much lower densities found here have potentially important 
implications for the predicted rate of inspiral
events, as discussed in more detail in \S\ref{sec:EMRI}.

In models with initial core radii $\lap 1.5$ pc (small enough to 
reproduce the currently-observed core), the density
of stellar BHs after 10 Gyr becomes large enough that self-interactions
between the BHs would be significant. A high enough density of BHs
would also tend to accelerate the relaxation of the stellar component.
While beyond the scope of this paper, tests of these predictions 
could be carried out via two-component, $f(E,L,t)$ models 
or $N$-body integrations.

\subsubsection {Summary}

The time scale for inspiral of $10\msun$ BHs to the center of
the Galaxy is longer in models with a core than in models with a cusp,
both because of the lower density of stars in the core, but also
because the dynamical friction force drops essentially to zero
at energies near the phase-space truncation energy that defines the 
core.
Assuming that the BHs and the stars have the same distributions
initially, the radius of peak density of the BHs tends to ``follow'' 
the stellar core as the latter shrinks.
The density of BHs can remain substantially less  than
that of the stars at  small radii ($r\lap 0.01-0.1$ pc)
even after the stellar core has shrunk to its observed size
of $\sim 0.5$ pc.

\section{Discussion}
\label{sec:discussion}

Topics discussed in this section include 
the dynamical consequences of ongoing star formation;
predicted rates of stellar tidal disruptions and gravitational
wave inspirals at the Galactic center; 
mechanisms for enhanced relaxation;
the effect of a parsec-scale stellar core on inspiral of intermediate 
mass black holes; 
the rate of production of hypervelocity stars by collisions
involving stellar BHs; and the connection between the core 
at the center of our Galaxy and the cores observed in other galaxies.

\subsection{Star formation}
\label{sec:starformation}
The Milky Way nuclear star cluster (NSC) 
sits at the center of a kiloparsec-scale
Galactic bulge or bar which consists mainly of old ($\sim 10$ Gyr)
evolved stars.
On sub-kpc scales, the Milky Way shows evidence for stellar populations 
with a range of ages.
Serabyn \& Morris (1996) argued that the conditions of the 
interstellar medium in the ``Central Molecular Zone'' would
lead inexorably to inflow of molecular material and to continuous
star formation activity in the central $\sim 10^2$ pc.
Figer et al. (2004) modelled the luminosity function of stars in the
inner $\sim 50$ pc and argued that single-burst star formation models could
be securely ruled out; they inferred a nearly constant rate of star
formation rate over the last $\sim 10$ Gyr.
These and other studies 
\citep[e.g.][]{Mezger:99,Philipp:99}
suggest that the NSC is not a simple inward extrapolation of the old bulge, 
but rather
consists of an intermediate age population that has been undergoing
continuous star formation since the creation of the Galaxy.

In a general way, continuous star formation strengthens the
picture presented here of a nuclear cluster that is less than $\sim$
one relaxation time old, by reducing the mean
age of stars from $\sim 10$ Gyr to $\sim 5$ Gyr.

Of more direct interest is the evidence for recent star formation
in the inner $\sim 0.5$ pc  \citep{Paumard:08}, 
the same region where the old stars exhibit a low-density core.
The total mass of the young stars currently observed in this region is 
probably less than
$10^4\msun$ \citep{Figer:08}, making them dynamically insignificant.
On the other hand, the starburst that created this population
may have been just the most recent
instance of an ongoing or episodic process.
How would such a ``source term'' modify the evolutionary 
calculations presented above?

The answer clearly depends on the accumulated stellar mass 
and on the radial dependence of the star formation rate;
both are highly uncertain.
Here we limit ourselves to answering a simpler question: How would a 
population of stars, formed initially in a disk, evolve 
against the background of a pre-existing stellar core?
We specifically ignore interactions between the disk stars, and
assume that the background stars are fixed in their distribution.
More detailed calculations, in which both populations are allowed
to evolve, will be described in a subsequent paper.

If the surface density of  the young stellar disk
is $\Sigma(r) \propto r^{-n}$,
the distribution of orbital energies is
$dN/dE = (dN/dr)(dr/dE)\propto |E|^{n-3}$.
To simplify the evolutionary calculation, 
we suppose that the orbits of the young stars ``randomize''
in orientation and eccentricity on a time scale shorter than $t_r$.
(Without trying to justify that assumption quantitatively, 
we note several mechanisms that might achieve this: standard relaxation as
in \S 3; ``resonant relaxation'' as in \S6.4; torques from another disk
or from a large-scale non-axisymmetric bulge component etc.)
The young stars can then be described by an isotropic 
$f$ where $f(E,t=0)\propto p(E)^{-1} N(E)\propto |E|^{n-1/2}$.
The corresponding space density is $\rho(r,t=0)\propto r^{-1-n}$.
If the disk is truncated at an inner radius $r_\mathrm{in}$,
then $\rho(r,t=0)\propto r^{-1/2}$ inside $\sim r_\mathrm{in}$.
We assume $n=2$ and that the disk stars are initially distributed
between $0.1$ pc and $0.4$ pc.

For the old stars, we assume a (fixed) 
$f_\mathrm{old}(E) = f_0 |E|^A$
which corresponds to a density
$\rho_\mathrm{old}(r) = \rho_0 r^{-\gamma}, \ \ \gamma = A+3/2, \ \ r < \rh$.
Substituting this expression for 
$f_\mathrm{old}$ into the energy-space diffusion 
coefficients $D_E, D_{EE}$ of equation~(\ref{eq:fp})
gives
\begin{subequations}
\begin{eqnarray}
D_E(E) &=& -\frac{32\sqrt{2}\pi^5}{3-2A} f_0 G^5\mh^3m^2\ln\Lambda |E|^{A-3/2}, \\
D_{EE}(E) &=& \frac{32\sqrt{2}\pi^5}{(1+A)(1-2A)} f_0 G^5\mh^3m^2\ln\Lambda |E|^{A-1/2}
\end{eqnarray}
\end{subequations}
again assuming that the gravitational potential is due to the SMBH alone.

We adopt units for time and energy such that
\beq
4\pi\Gamma f_0 [T][E]^A = 1
\eeq
where $\Gamma \equiv 4\pi G^2m^2\ln\Lambda$.
A natural energy unit is
\beq
[E] = \psi_0 \equiv \frac{G\mh}{\rh}
\eeq
which makes the unit of time
\beq
[T]^{-1} = 16\pi^2G^2m^2\ln\Lambda f_0 \psi_0^A.
\eeq
In dimensionless variables,
equation~(\ref{eq:fp}) for the young stars then becomes 
\begin{subequations}
\begin{eqnarray}
\frac{\partial f}{\partial t} &=& - E^{5/2}\frac{d{F}}{d E},\\
F(E) &=& -\frac{2}{(1+A)(1-2A)} E^C \frac{d}{d E} 
\left(E^{B}f\right) 
\end{eqnarray}
\label{eq:fp7}
\end{subequations}
where
\beq
B = \frac{-(1+A)(1-2A)}{3-2A}, \ \
C = \frac{-1+6A-8A^2}{2(3-2A)}.
\eeq

The steady-state solution is given by setting
$F(E)=0$:
\beq
f(E,t\rightarrow\infty) \propto E^{-B}.
\eeq
Choosing $A=1/4$ ($\gamma=7/4$) implies $B=-1/4$; in other words,
test particles interacting with stars in a Bahcall-Wolf cusp
evolve also to the Bahcall-Wolf form.
However, other choices for $A$ 
imply different steady-state solutions for the young stars.

\begin{figure}
\includegraphics[width=0.50\columnwidth,angle=-90.]{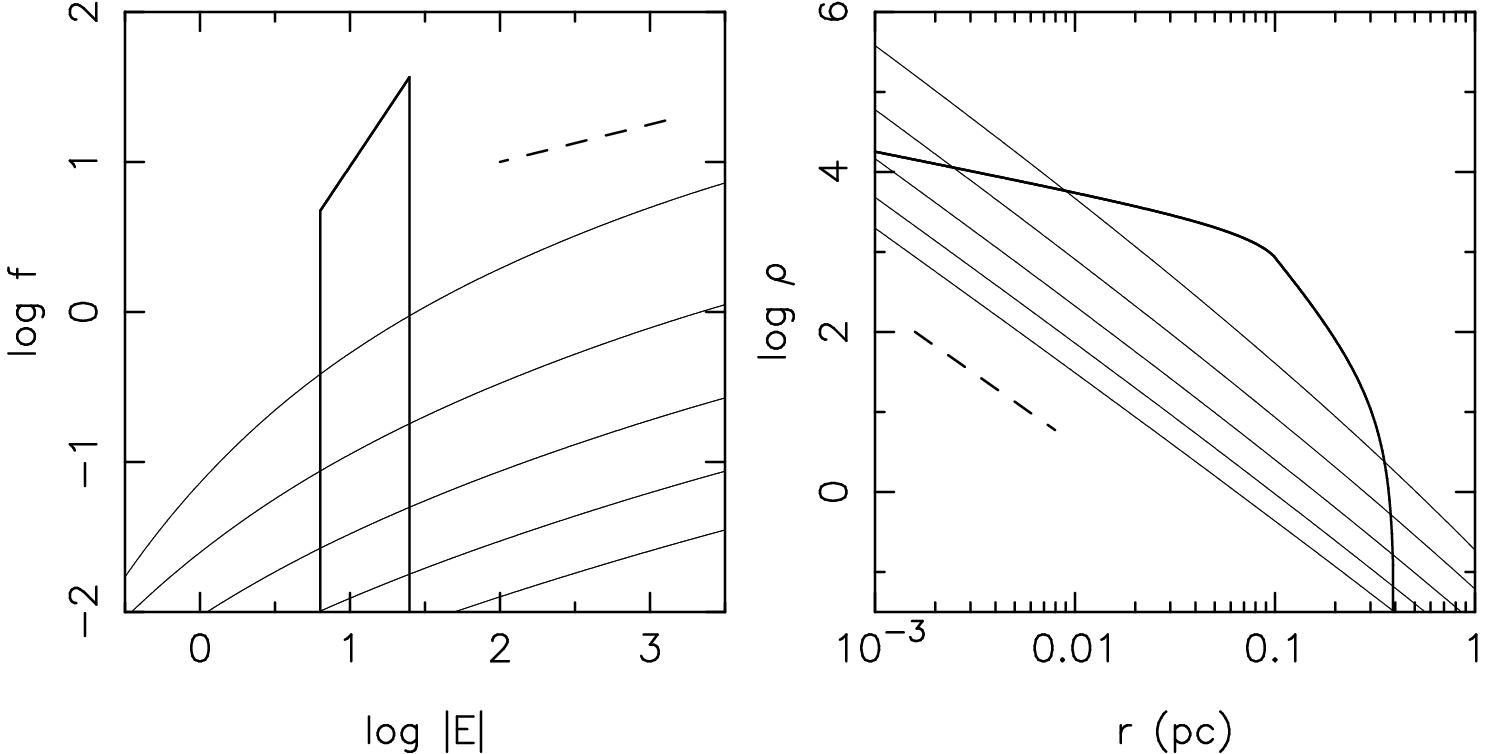}
\includegraphics[width=0.50\columnwidth,angle=-90.]{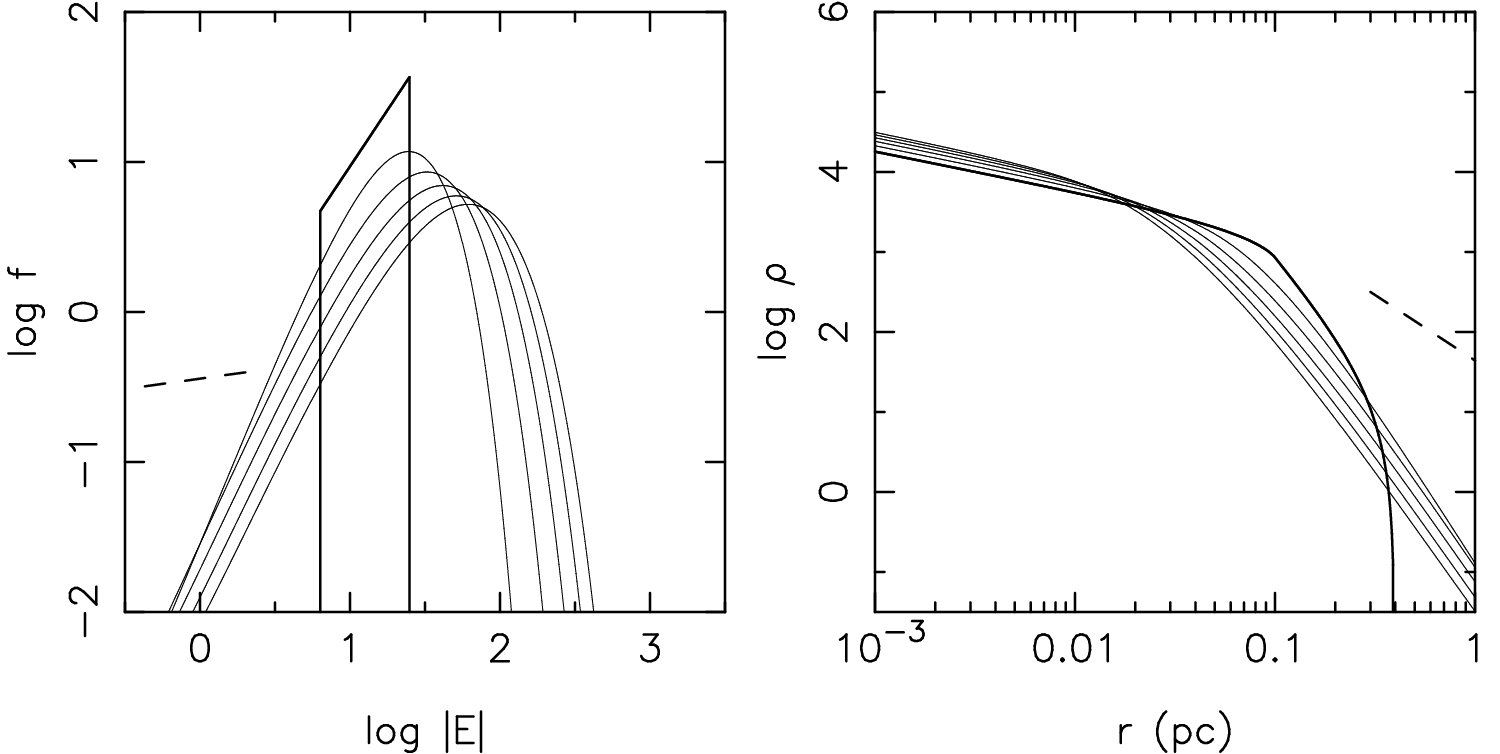}
\caption{\label{fig:1} Evolution of $f(E,t)$ and $\rho(r,t)$ describing 
an isotropic population of ``young'' stars that formed initially in a disk,
and subsequently scatter off of the ``old'' stars, assumed to have a fixed
phase-space distribution.
Initial conditions are shown in bold; subsequent times are
$(1,2,3,4,5)\times 10^9$ yr, assuming that the mass in the old
population within $1$ pc is $1.5\times 10^6\msun$.
Length units were scaled to the Galactic center.
(a) $f_\mathrm{old}\propto E^{0.25}$,
$\rho_\mathrm{old}\propto r^{-7/4}$,
the Bahcall-Wolf form.
(b) $f_\mathrm{old} \propto E^{-3/4}$, 
$\rho_\mathrm{old}\propto r^{-3/4}$.
Dashed lines show the steady-state slopes.
\label{fig:starform}}
\end{figure}

Figure~\ref{fig:starform} shows time-dependent
solutions to equation~(\ref{eq:fp7})
under two assumptions about the 
 background population.
In both cases, the (fixed) density of the old population was normalized
to give a mass within one parsec of $1.5\times 10^6\msun$.
Figure~\ref{fig:starform}a assumes
$\rho_\mathrm{old}\propto r^{-7/4}$ ($A=1/4$),
a Bahcall-Wolf cusp.
In this case, the relaxation time at the initial disk radius
is just a few Gyr, and the young stellar population also
reaches a distribution close to the Bahcall-Wolf form after
$5$ Gyr, as expected.
Gravitational encounters with the old stars tend to redistribute 
the young stars to both smaller and larger radii.

Figure~\ref{fig:starform}b shows the case
$\rho_\mathrm{old}\propto r^{-3/4}$ ($A=-3/4$),
a lower-density core.
Now, the relaxation time increases toward the center;
the distribution of young stars hardly changes inside
the original inner disk radius, and the net result of gravitational
encounters is mostly to scatter the disk stars to larger
radii.
The steady-state form, $\rho\propto r^{-1.63}$, is only gradually
approached, and only at radii outside the initial disk radius.
There is essentially no evolution toward a Bahcall-Wolf cusp.

In both cases, the number of young stars that remain within
the original outer disk radius after 5 Gyr is a small fraction
of the initial number: $\sim 10^{-3}$ in the $A=1/4$
case and $\sim 0.15$ for $A=-3/4$.

These results, while very preliminary, suggest that ongoing star
formation need not greatly modify  the conclusions that were
arrived at above.
As long as star formation occurs against the backdrop of a stellar core,
the density profile of the young stars can remain relatively
flat inside the initial disk radius.
Their mean density is also strongly diluted by encounters with
the older stars.

This conclusion may need to be modified in the case of the stellar remnants.
The $\sim10-100$ very massive blue giant stars in the
stellar disks will probably end their lives as an equal number of
$10\msun$ BHs, with total mass $\sim10^2-10^3\msun$.
If such star forming events occur once per $10^8$ yr, then 
$10^4\msun-10^5\msun$ in stellar BHs could accumulate in the central
parsec over $10$ Gyr.
\footnote{I thank Tal Alexander for pointing this out.}
The expected number of remnants will also depend strongly on
the form of the initial mass function (IMF); for instance, a ``top-heavy''
IMF \citep{Paumard:06} would produce many more stellar BHs per unit 
of total mass.

\subsection{Rates of tidal disruption}

The existence of a core implies a smaller rate of stellar tidal
disruptions at the Galactic center than in models that assume
a density cusp.
The rate of scattering of stars into the SMBH's tidal disruption sphere,
$r\le r_t$, can be written
\beq
\dot N = \int {\cal F}(E) dE
\label{eq:dotN}
\eeq
where ${\cal F}(E)$ is the number of stars scattered per unit
time and unit energy into $r_t$.
In the models considered here, setting the core radius to zero
gives $\rho\propto r^{-1.8}$ at all radii (eq.~\ref{eq:rho}),
and this is also approximately the form of the density in the
single-mass Bahcall-Wolf steady-state solution at $r\lap\rh$,
$\rho\propto r^{-1.75}$.
Both density laws, in turn, are close to a singular isothermal sphere, 
$\rho\propto r^{-2}$, for which the feeding rate has
been shown to be
\beq
\dot N \approx 4.6\times 10^{-4} {\rm yr}^{-1} 
\left({\sigma\over 90\ {\rm km\ s}^{-1}}\right)^{7/2}
\left({\mh\over 4\times 10^6\msun}\right)^{-1}
\label{eq:dotN2}
\eeq
assuming Solar-type stars \citep{WM:04}.

Carving out a core changes $\dot N$ for two reasons.
(1) The energy integral, equation~(\ref{eq:dotN}), 
now has $E_b$ as a lower limit rather than $-\infty$.
(2) The diffusion coefficients $D_{\cal R}, D_{\cal RR}$ 
that determine the scattering rate at every $E$ 
(eqs.~\ref{eq:DR}) are smaller due again to the 
absence of stars with $E<E_b$.

A commonly made approximation \citep[e.g.][]{CK:78}
is to ignore the contribution to $F(E)$ from scattering
off of stars with energies that are smaller (i.e. more
tightly bound) than $E$.
In this approximation, the diffusion coefficients are not
changed, at energies $E>E_b$, by the truncation of $f$ at
$E_b$, and the only change in $\dot N$ comes from the change
in the lower integration limit for equation~(\ref{eq:dotN}).

\begin{deluxetable}{lr}
\tablecaption{Stellar tidal disruption rates \label{tbl-1}}
\tablewidth{0pt}
\tablehead{
\colhead{$r_b$ (pc)} & \colhead{$\dot N$ (yr$^{-1}$)}}
\startdata
0   & $4.6\times 10^{-4}$ \\
0.1 & $2.7\times 10^{-4}$ \\
0.2 & $1.7\times 10^{-4}$ \\
0.5 & $5.6\times 10^{-5}$ \\
1   & $2.7\times 10^{-5}$ \\
2   & $2.0\times 10^{-5}$ \\
\enddata
\end{deluxetable}

Table 1 shows $\dot N$ for the Milky Way computed under this approximation,
for various values of $r_b$, assuming an ``isotropic core.''
(Recall that $r_b$ is essentially equal to the core radius
of the corresponding density profile, Figure~\ref{fig:core1},
so that setting  $r_b\approx 0.5$ pc gives a core of roughly 
the correct size for the Milky Way.)
A 0.5 pc core implies a tidal flaring rate that is almost
an order of magnitude smaller than for a coreless cusp.

This conclusion should be considered extreme since it ignores changes
in $f$ that will necessarily tend to refill the depleted orbits
over relaxation time scales.
In principle, the time integrations of $f$ in \S 4 could be used
to compute the evolution of the tidal disruption rate.
We postpone that calculation to a later paper, but make a 
related point here: if $f$ is not in a steady state, the standard
expressions for the flux into the loss cone (like the expressions that
were used to derive eq.~\ref{eq:dotN2}) are not strictly valid
\citep{MM:03,MW:05}.
This highlights the need to develop a more complete theory of
time-dependent loss cones.

\subsection{Enhanced relaxation}

The evolution equations for $f$ and $f_{\rm BH}$
that were solved in \S 4 and \S 5
were based on a standard, orbit-averaged Fokker-Planck treatment
of gravitational encounters.
The results were scaled to physical time units assuming that
relaxation is driven by perturbers with masses of roughly $\msun$.

Both of these assumptions have been questioned, in recent papers
that argue for more efficient relaxation near the Galactic center
\citep{HA:06b,Perets:07}.
Two  important points have been made:
(1) The effectiveness of gravitational encounters at inducing changes
in orbital angular momenta is increased for stars orbiting within the 
gravitational field of a SMBH, $r<\rh$, since the orbits are nearly Keplerian
and they maintain their orientations for many radial periods, allowing
torques to accumulate linearly with time 
(``resonant relaxation''; Rauch \& Tremaine 1996).
(2) If there is a distribution of masses in the scattering
population, the effective relaxation time is determined
by the second moment of the mass function (eqs.~\ref{eq:both}).
The presence of even a small population of very massive objects
(``massive perturbers'') in the region of interest can reduce the effective
relaxation time considerably.

We briefly discuss the applicability of these ideas to the
evolutionary models discussed here.

{\it Resonant relaxation:} Resonant relaxation 
is relevant to the timescale for isotropization that was
computed in \S 4.
The resonant relaxation time is
\beq
t_{\rm RR}\approx {1\over N(<a)} \left({\mh\over m_\star}\right)^2 
{P^2(a)\over t_{\rm precess}}
\eeq
\citep{HA:06b}
where $N(<a)$ is the number of perturbing stars, of mass $m_\star$, inside
the orbit of the test star whose semi-major axis is $a$; $P$ is
the orbital period; and $t_{\rm precess}$ is the time scale above which
orbits lose their coherence due to precession.
If precession is due primarily to the gravitational force from the $N$ stars,
distributed spherically inside $r=a$, 
then
\beq
t_{\rm precess}\approx {1\over N(<a)} {\mh\over m_\star} P(a)
\eeq
and
\begin{subequations}
\begin{eqnarray}
t_{\rm RR} &\approx& {\mh\over m_\star} P(a) \\
&\approx& 2\times 10^{11} {\rm yr} \left({\mh/m_\star\over 4\times 10^6}\right)
\left({a\over 1\ {\rm pc}}\right)^{3/2},
\label{eq:trr}
\end{eqnarray}
\end{subequations}
{\it independent} of the density of perturbers.
Hopman and Alexander (2006b) note that this time 
falls below the standard relaxation time (\ref{eq:tr_spitz})
at a distance $0.1-0.5$ pc from \sgr.
This suggests  that resonant relaxation may reduce somewhat
the time scale for isotropization compared with the values 
computed in \S 4.

The counter-intuitive result that $t_{\rm RR}$ is independent of
the density of perturbing stars (eq.~\ref{eq:trr})
is due to the fact that the coherence time $t_{\rm precess}$ becomes
long as $N$ becomes small, allowing even small torques to build up 
for long times.
In the core models discussed here, $N$ can be essentially zero,
and other processes would likely begin to dominate the
precession rate.
For instance, a nuclear bar \citep[e.g.][]{Alard:01}
would generate a non-axisymmetric component to the gravitational 
potential in its interior, setting an upper limit
to the precession time and reducing the effectiveness of resonant
relaxation compared with the expressions given above as $N\rightarrow 0$.

{\it Massive perturbers:} 
Equation~(\ref{eq:both}) says that the rate of gravitational 
scattering by a background population of perturbers scales as
\beq
n_p \langle m_p^2\rangle
\eeq
where $n_p$ is the number density of perturbers of individual 
mass $m_p$; the brackets denote a number-weighted average.
This expression ignores differences in the velocity distribution
between the different populations, which is reasonable if the
relaxation time is long and/or if the perturbers were recently formed.
Perets et al. (2007) noted that the mass spectrum of giant molecular
clouds implies that they should dominate the scattering rate beyond 
a few parsecs from \sgr, reducing the relaxation time in this region
by as much as several orders of magnitude.
Inside $\sim 5$ pc, they suggested that gas clumps in the circumnuclear
gas disk, with masses $10^3-10^5\msun$, might be similarly important.

Massive perturbers beyond $\sim 1$ pc would affect the 
distribution of stars near the SMBH in two distinct ways
\citep{Perets:07}.
1. Deflection of unbound (with respect to the SMBH) stars onto
radial orbits would fill in some of the phase space volume that
was evacuated by formation of the core. 
This population would have a spatial distribution 
$n\sim r^{-1/2}$, similar to that of the pre-existing core.
2. Three-body interactions of field binaries deflected by massive perturbers 
can create a population of bound stars around the SMBH, 
initially on very eccentric
orbits.
The radial distribution of these bound stars will reflect the semi-major
axis distribution of the parent binary population, which is
uncertain.
The capture rate is estimated to be as large as 
$\sim 10^{-4}$ yr$^{-1}$ in the inner parsec;
the accumulated mass could therefore potentially exceed 
the number of stars in the evolutionary models considered here.
These arguments, while very approximate, suggest that massive
perturbers could compete with stellar-mass perturbers in terms
of refilling an evacuated core.

In a general way, the fact that the Milky Way {\it does} contain a
low-density core implies an upper limit on the effectiveness of any
relaxation process, particularly those that change orbital energies.

\subsection{Extreme-mass-ratio inspirals}
\label{sec:EMRI}
Inspiral of compact remnants (stellar mass BHs, neutron stars, white dwarves)
into a SMBH is accompanied by the emission of gravitational waves
with frequencies that will be detectable by the Laser Interferometer
Space Antenna (LISA) \citep{SR:97,Barack:04,Amaro:07}.
Event rates of these extreme-mass-ratio  inspirals (EMRIs) are generally
computed assuming that the stars and stellar remnants are distributed
in a relaxed, multi-mass density cusp \citep{HB:95,HA:05,Hopman:09}.
In such a cusp, the density of solar-mass stars follows
$n\sim r^{-1.5}$ while the $10\msun$ BHs have a steeper dependence,
$n\sim r^{-1.75}-r^{-2}$.
The radius at which the mass density of BHs rises above that 
of the less massive objects depends on the choices made for the mass function,
and (in the time-dependent models) the initial spatial distributions and the 
elapsed time.
Typically, $\rho_{\rm BH}>\rho_\star$ inside $0.01-0.1$ pc
\citep{Freitag:06a,HA:06a}.
The EMRI event rate is dominated by BHs (as opposed to neutron
stars or white dwarves) due to their high masses and high mass densities.
Most of the signal is contributed by BHs inside $\sim 10^{-2}$ pc
(e. g. Hopman \& Alexander 2006a, Figure~2).

Here we estimate BH inspiral rates for models of the nuclear star
cluster that include a core.
A distribution-function based approach, similar to what was used
above to compute stellar tidal disruption rates, would require an 
additional Monte Carlo calculation to estimate the probability that 
a star on a loss cone orbit will evade being scattered directly into 
the SMBH before emitting gravitational waves.
Instead, we follow the more approximate treatments in 
Hils \& Bender (1995), Hopman \& Alexander (2005) and Ivanov (2002)
based only on the density profiles of the two components.
We first derive an expression for the inspiral rate based on the 
observed stellar distribution, Figures 1 and 3.
Event rates predicted by the evolving, two-component models of \S 5
are then computed.

Gravitational wave emission is dominated by BHs that are scattered into
the SMBH from orbits with $a\lap a_{\rm crit}$, where $a_{\rm crit}$  is the
orbital radius such that the decay time due to emission of
gravitational waves equals the time for the BH to be scattered
in or out of the loss cone by stars or by other BHs.
The angular momentum of a loss cone orbit is
\beq
J_{\rm lc} \approx {4G\mh\over c}.
\eeq
Stars with $a\lap r_{\rm crit}$ can avoid scattering for a time
long enough to spiral in via emission of gravitational waves.
The inspiral rate is approximately
\beq
\dot N_{\rm GW}\approx \int_0^{a_{\rm crit}}
{f_{\rm BH}(a) N(a) \over t_r(a) \ln\Theta} da
\label{eq:ndotGW}
\eeq
where $N(a)da$ is the number of stars and BHs with semi-major axes
$a$ to $a+da$,
$f_{\rm BH}\le 1$ is the fraction of objects which are BHs, 
and $\Theta = J_c/J_{\rm lc}$.

The gravitational wave inspiral time for an eccentric orbit,
$J\ll J_c$, is
\citep{HA:05}
\beq
t_{\rm GW} = {3\times 2^{14}\over 85} {\sqrt{G\mh a}\over c^2} 
{\mh\over \mbh} \left({J\over J_{\rm lc}}\right)^7
\label{eq:tGW}
\eeq
where $\mbh$ is the mass of a stellar BH.
The critical radius is defined as the radius where
$t_{\rm GW}=t_{\rm lc}$ with
\beq
t_{\rm lc} = \left({J_{\rm lc}\over J_c}\right)^2 t_r ,
\label{eq:tlc}
\eeq
the diffusion time into the loss cone.
Combining (\ref{eq:tGW}) and (\ref{eq:tlc}), and identifying the relaxation
time on an orbit with its value at $r=a$, gives an implicit relation
for $a_{\rm crit}$:
\beq
{85\over 3} {\mbh\over\mh} \left({G\mh\over a_{\rm crit}^3}\right)^{1/2}
t_r(a_{\rm crit}) = 2^{10}
\label{eq:sacrit}
\eeq
the solution to which determines the upper limit to the rate integral
(\ref{eq:ndotGW}).

For the relaxation time, we use the estimate plotted in
Figure~\ref{fig:tr} for the Galactic center, which was
based on the parametric model fit to the number counts, 
Figure~\ref{fig:fits}, 
with $n\propto r^{-1/2}$ enforced at small radii.
This is the flattest central dependence consistent with an isotropic
$f$ for the stars (\S 3), and is also similar to what is found in the
evolving models (\S 4, 5).
The relaxation time plotted in Figure~\ref{fig:tr} can be written
\beq
t_r(r) \approx 1\times 10^{10} {\rm yr} \left({\tilde m\over\msun}\right)^{-1}
\left({r\over 0.2 {\rm pc}}\right)^{-1}, \ \ \ \ r\lap 0.2\ {\rm pc}
\label{eq:trnew}
\eeq
where $\tilde m$ is defined in equation~(\ref{eq:defmtilde}) 
and accounts for the possibility that stellar BHs
may contribute significantly to the density of scatterers.
(Note however that this expression assumes a particular normalization
for  the {\it total} mass density; this assumption will be relaxed 
below).
In our two-component model, $\tilde m$ is fixed by $f_{\rm BH}$:
\beq
{\tilde m\over m} = {1-f_{\rm BH}+f_{\rm BH}\mbh^2/m^2\over 
1-f_{\rm BH}+f_{\rm BH}\mbh/m} \approx {1+99f_{\rm BH}\over 1+9f_{\rm BH}}
\eeq
with $m$ the stellar mass; the last relation assumes 
$\mbh=10 m$.

Substituting (\ref{eq:trnew}) into (\ref{eq:sacrit}) gives 
for the critical radius
\beq
a_{\rm crit} \approx 0.08\ {\rm pc} \left({\tilde m\over\mbh}\right)^{-0.4}.
\eeq
For $\tilde m/\mbh = 0.1 (0.5) 1$,
$a_{\rm crit}/{\rm pc} = 0.2 (0.11) 0.08$.
These are somewhat larger than the critical radii computed
assuming a relaxed density cusp ($a_{\rm crit}\approx 10^{-2}$ pc)
but are still small compared with the observed core radius of $\sim 0.5$ pc.

The event rate (\ref{eq:ndotGW}) becomes
\beq
\dot N_{\rm GW} \approx 3\times 10^{2} {\rm Gyr}^{-1} f_{\rm BH}
\left({\tilde m\over\mbh}\right)^{-0.4} {N_{0.1}/10^4\over
\ln\Theta/5}
\label{eq:dotNGW}
\eeq
where $N_{0.1}$ is the number of stars and BHs inside $0.1$ pc;
the density normalization assumed in making Figure~\ref{fig:tr},
equation~(\ref{eq:rho2}), implies $N_{0.1}\approx 8.6\times 10^3$.

Setting $f_{\rm BH} = 0.001$ and $\tilde m\approx 1\msun$
in this formula gives an estimate of the inspiral rate 
for an unsegregated model in which
the BHs follow the same density profile as the observed stars.
Not surprisingly, the resultant rate is very low, 
$\dot N\approx 1$ Gyr$^{-1}$. 
If instead $f_{\rm BH}$ is set to the higher values found in the
steady-state models, i.e. $f_{\rm BH} = 0.01 (0.05) 0.1$,
equation~(\ref{eq:dotNGW}) gives 
$\dot N_{\rm GW} \approx 2 (9) 15 $ Gyr$^{-1}$.
These rates are still 1-2 orders of magnitude lower than those
in the steady-state models, due to the very different {\it total}
densities assumed at the relevant radii; 
for instance, Hopman \& Alexander (2006a)
find $\dot N_{\rm GW}\approx 300$ Gyr$^{-1}$ in models that have
$0.01\lap f_{\rm BH}\lap 0.1$ at $r\approx a_{\rm crit}$.
The magnitude of the difference is at first sight surprisingly small
given that
the steady-state models have total densities at $10^{-3}-10^{-2}$ pc that
are orders of magnitude higher than here.
The reason is that the longer relaxation time in the core models
implies a $\sim 10\times$ larger value of $a_{\rm crit}$
and a $\sim 10^3\times$ larger volume from which inspirals can occur 
\citep[e.g.][]{HA:05}.

\begin{figure}
\includegraphics[clip,width=0.425\textwidth]{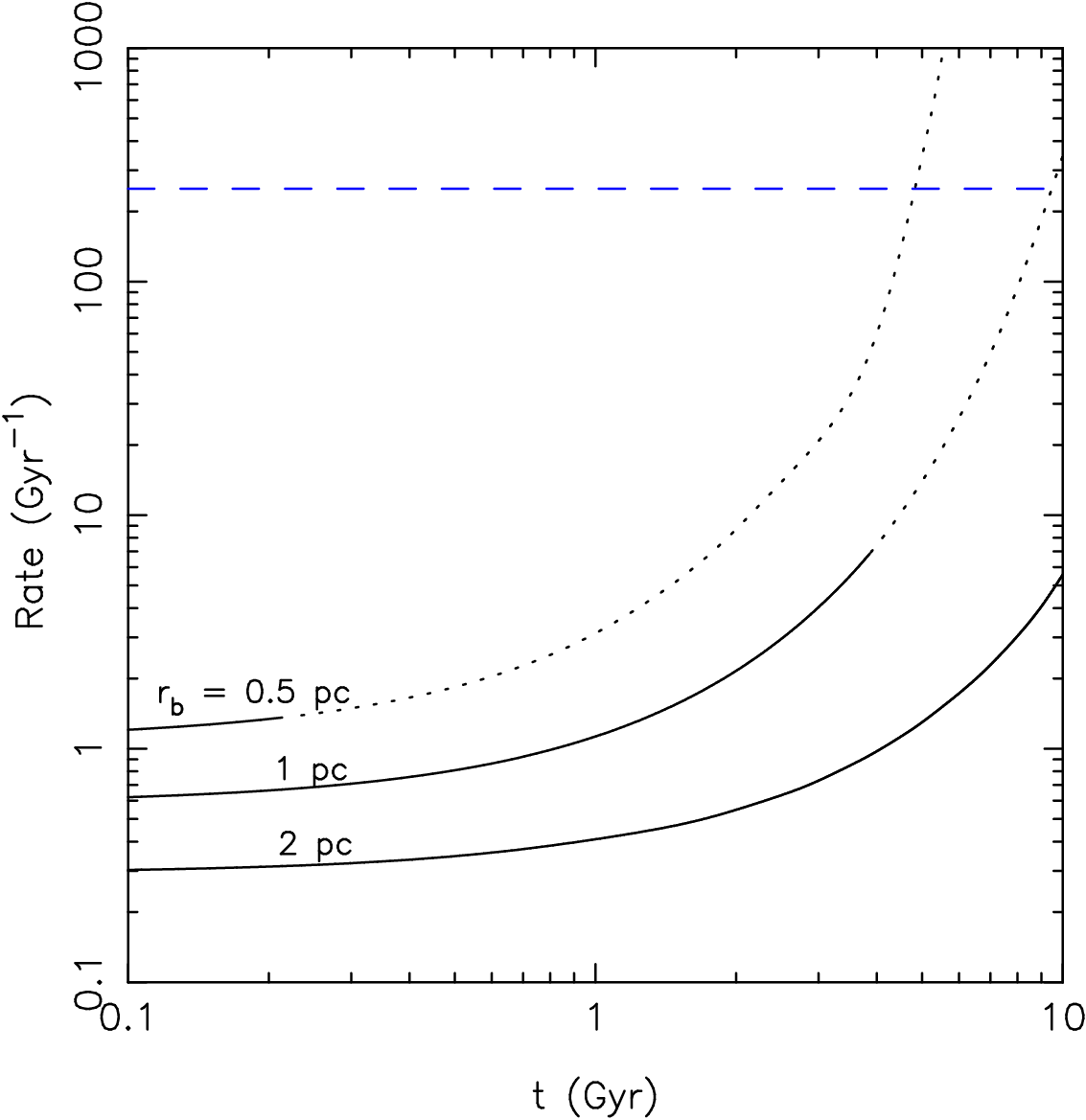}
\caption{Evolution of the BH inspiral rate in the models
of Figure~\ref{fig:rhorcore}.
The solid part of each curve terminates at the time when  
the stellar core radius is 0.5 pc.
Dashed (blue) line shows the event rate computed by Hopman
\& Alexander (2006a) in steady-state, mass-segregated models
(cf. their Fig. 2).
}
\label{fig:GWs}
\end{figure}

In the time-dependent models of \S 5, not only are total densities
smaller than in the steady-state models, but $f_{\rm BH}$ also remains 
well below its value in a mass-segregated cusp until late times.
Figure~\ref{fig:GWs} shows BH inspiral rates computed in the
same way as above, using the densities of stars and BHs in
the time-dependent models.
In these models, the BHs have initially the same spatial
distribution as the stars; as the stellar core shrinks,
the BHs ``follow'' it inward, increasing both their density
near the center and their relative density with respect to the
stars (Figure~\ref{fig:rhorcore}).
By the time the stellar core shrinkss to its observed value
of $\sim 0.5$ pc, the BH density at $r\lap a_{\rm crit}$ is
still much lower than in the steady-state models and
$\dot N_{\rm GW}$ is also correspondingly lower.

It must be emphasized that these evolutionary models
are idealized and that one could imagine other, reasonable
initial conditions that would produce rather different rates
after 10 Gyr.
For instance, the distribution of BHs might have exhibited
some degree of segregation even at early times.
The main point to be made here is that in models for the
nuclear star cluster that contain a core, $\dot N_{\rm GW}$
could plausibly be as much as 10-100 times lower than in
models based on a steady-state, mass-segregated density cusp.

\subsection{Inspiral of an IMBH}
\bigskip
Intermediate mass black holes (IMBHs), with masses of $10^2-10^3\msun$,
may form in dense star clusters through runaway mergers of massive
stars \citep{PZM:02,Freitag:06b}.
Inspiral of an IMBH into
the Galactic center is usually modelled assuming that the dynamical
friction force originates in a stellar density cusp with 
$\rho\sim r^{-\gamma}, 1.5\lap\gamma\lap 2$ \citep{BGZ:06,Matsu:08,LB:08}.
In these circumstances, inspiral continues until the IMBH 
reaches a distance from \sgr\ such that the enclosed stellar mass
is roughly equal to $m_{\rm IMBH}$, or 
$10^{-3} {\rm pc}\lap r\lap 10^{-2} {\rm pc}$.

If there is a pre-exisiting core in the stellar distribution,
inspiral would stall at roughly $1/2$ the
core radius, or $\sim 0.25$ pc in the case of the Galactic center
(Figure~\ref{fig:inspiral}),
independent of the IMBH mass.
Furthermore, as noted above, the orbital eccentricity of the
IMBH would not be expected to decrease strongly during the inspiral.

Merritt et al. (2009) and Gualandris \& Merritt (2009) noted one 
consequence of a ``stalled'' IMBH on an eccentric orbit: the IMBH 
very efficiently randomizes the orbits of ambient stars, producing
a nearly ``thermal'' distribution of orbital eccentricities,
$N(<e)\sim e^2$, on Myr time scales.
Merritt et al. (2009) postulated stalling radii inside $\sim 0.1$ pc
in order to explain the observed distribution of the S-star orbits
\citep{Ghez:08,Gillessen:09}.
The somewhat larger stalling radii made plausible here
suggest that IMBHs might be lurking on somewhat wider orbits,
roughly the size of the two stellar disks at $0.1 {\rm pc}\lap r\lap 0.5$ pc
\citep{Paumard:06,Lu:09,Bartko:09a}.
As shown by Levin et al. (2005) and Berukoff \& Hansen (2006),
an IMBH at these radii could play a role in truncating the stellar disks
and scattering disk stars onto inclined and eccentric orbits.
Those authors assumed efficient inspiral of the IMBHs, which
limited the time over which interactions could occur. 
If orbital decay stalls at distances of $\sim 0.2$ pc, 
dynamical interactions with disk stars could
be prolonged indefinitely, potentially resulting in much 
greater changes in the stellar orbits.

An inspiralling IMBH also ejects stars and stellar remnants
via three-body interactions with the IMBH/SMBH binary.
Some of these stars receive kicks greater than $\sim 10^3$ km s$^{-1}$,
allowing them to escape into the Galactic halo
as hyper-velocity stars (HVSs) \citep{Yu:03}.
Simulations of this process \citep{Levin:06,BGZ:06,LB:08} also 
typically assume a steep density cusp for the stars.
The resulting rate of ejection of HVSs increases rapidly with
time as the IMBH spirals in, peaking when the IMBH reaches
its (small) stalling radius of $\sim 0.01$ pc, then falling
off due to the local depletion of stars.
The stellar density at this radius is assumed to be
$\gap 10^8\msun$ pc$^{-3}$ initially, roughly the density implied
by the (coreless) power-law model of equation~(\ref{eq:rho}). 

In the core models considered here, the rate of production of HVSs
would be much smaller than in the relaxed models due to the lower
stellar densities inside $\sim 1$ pc.
In addition, the IMBH would stall at a larger radius of $\sim 0.2$ pc.
The density at this radius is $\lap 10^6\msun$ pc$^{-3}$, 
resulting in a $\sim 10^2$ times smaller rate of escapers than the 
peak values obtained during infall in models with a steep cusp, 
i.e. $0.1-1$ Myr$^{-1}$.
On the other hand, at these low rates, ejections by the IMBH
would hardly affect the ambient stellar density and the production
of HVSs could continue indefinitely at an approximately constant
rate.
Production of $\sim 10$ HVSs would therefore require a span of 
$\sim 10-100$ Myr; in fact the observed span of travel times for
escaping HVSs is $\sim 2\times 10^8$ yr \citep{Brown:07}, roughly
consistent with this crude estimate.

\subsection{Dynamical interactions that postulate a high density
of stellar black holes}
A dense cluster of stellar-mass BHs has been invoked as a potential
solution to a number of other problems of collisional dynamics at the Galactic
center.
Examples include: (1) removing stellar envelopes via physical collisions between BHs and 
stars \citep[e.g.][]{Dale:09}; 
(2) randomizing the orbits of young stars via gravitational scattering 
off of BHs \citep[e.g.][]{Perets:09};
(3) production of HVSs through encounters with BHs \citep[e.g.][]{OL:08}.
Typically $n_{\rm bh}\sim r^{-2}$ is assumed,
as in the relaxed, mass-segregated models, implying a mass in BHs
of $\sim 10^4\msun$ within one 0.1 pc.
The $10-100$ times lower BH densities found in some of the
evolutionary models presented here would imply correspondingly
lower rates of interaction.

We note that alternative mechanisms exist for solving many of these
outstanding problems.
For instance, a single IMBH can randomize the orbits
of young stars even more efficiently than a BH cusp \citep{MGM:09}.

\subsection{Cores and nuclear star clusters}

As shown above (\S 4), the relaxation time at the Galactic center 
is short enough that a parsec-scale core will shrink appreciably 
over the course of 10 Gyr.
What is the connection between such a hypothesized initial core,
and the cores that are observed in spheroids brighter than
$M_B\approx -19.5$ \citep{Cote:07}?

The classification
of spheroids into cored or coreless families is based on data with
an angular resolution of $\sim 0.1''$, corresponding to a linear
size of $\sim 1$ pc at a distance of the Virgo galaxy cluster.
Neither the current core at the center of the Milky Way, nor
the larger initial core postulated here, would be easily
discernable at this distance.
Instead, the Milky Way would likely be classified as a galaxy with a 
nuclear star cluster (NSC): its luminosity profile
is relatively flat outside $\sim 10$ pc
and rises steeply inside  \citep{Launhardt:02,SME:08,Graham:09}.
The observed core sits atop that star cluster.
NGC 205 also has both a nuclear star cluster and a core;
the core radius is $\sim 0.12''\approx 0.5$ pc \citep{Merritt:09},
making it very similar to the Milky Way core
(although it is not clear that NGC 205 contains a SMBH; Valluri et al. 2005).

Many other galaxies with NSCs could also contain undetected cores.
Bright young stars, like those at the center of the
Milky Way, M31, and other galaxies with NSCs \citep{Marel:07},
would tend to mask the existence of a core in the old population,
as indeed they did until very recently at the Galactic center.

While most NSCs are too small for their internal structure to be
resolved, the half-mass relaxation time $t_{\rm nuc}$ can be 
reliably estimated for many,
and its mean dependence on host galaxy (not NSC) luminosity is
\beq
\log_{10} (t_{\rm nuc}/{\rm yr}) = 9.38 - 0.434\left(M_B+16\right)
\eeq
\citep{Merritt:09}
where $M_B$ is the absolute blue magnitude of the bulge component.
If we assume that galaxies with NSCs also contain SMBHs, and that the
relaxation time at $\rh$ is no greater than its value at the NSC
half-light radius, then relaxation times drop below $10$ Gyr
at $M_B\approx -17$, slightly fainter than the estimated
luminosity of the Milky Way bulge, $M_B=-17.6$
\citep{MH:03}.

These arguments suggest that cores comparable in size to SMBH
influence radii might exist in other galaxies with NSCs.
Relaxation times in the NSCs are expected to be short
enough that such cores could shrink appreciably in 10 Gyr.

\section{Conclusions}

1. The distribution of old stars at the Galactic center exhibits a low-density
core of radius $\sim 0.5$ pc. 
The deprojected central density is poorly constrained but is consistent with
zero.

2. Assuming that the old stars trace the mass in the inner parsec, 
the two-body relaxation time (for Solar-mass stars) is nowhere shorter
than $\sim 5$ Gyr.
The relaxation time at the influence radius of \sgr, $\rh\approx 2.5$ pc, 
is robustly estimated to be $20-30$ Gyr.

3. Reproducing the observed distribution of old stars with a steady-state 
distribution function requires a strongly truncated phase-space density at 
low energies and/or low angular momenta.
If the stellar density increases more slowly than $r^{-0.5}$ toward
\sgr, the velocity distribution must be anisotropic in the inner parsec,
with a deficit of eccentric orbits (``anisotropic core'').
Otherwise the distribution function can be isotropic (``isotropic core'').

4. Anisotropic core models evolve toward isotropy on a Gyr time scale.
In the process, the core radius decreases only slightly.
The observed (small) degree of anisotropy at the Galactic center  is consistent
with such models at both early and late times.

5. On a longer time scale, gravitational encounters produce changes in
stellar orbital energies, causing a pre-existing core to shrink.
Initial core radii in the range $1-1.5$ pc evolve, after $10$ Gyr,
to cores of the currently observed size.

6. The dynamical friction force acting on an inspiralling massive body
falls essentially to zero at roughly 1/2 the stellar core radius.
This results in an accumulation of $10\msun$ black holes in a shell
just inside the stellar core.
Orbital decay of an intermediate mass black hole would also be expected 
to stall at this radius, rather than the much smaller stalling radius
expected in a dense stellar cusp.

7. The expected density of $10\msun$ black holes in the 
inner parsec depends sensitively on their initial distribution
and on the elapsed time, but may be substantially lower than in models
that assume the absence of a stellar core. 
Rates of gravitational wave driven inspirals of stellar-mass black
holes are 1-2 orders of magnitude lower than predicted by steady-state 
models with a mass-segregated density cusp.

\acknowledgements

Don Figer, Sungsoo Kim, Rainer Sch\"odel, and especially Tal Alexander
read early versions of this paper and provided helpful comments and 
corrections.
Useful discussions with M. Messineo, H. Perets and E. Vassiliev 
are also acknowledged.
The author was supported by grants AST-0807910 (NSF) and NNX07AH15G (NASA).

\appendix
\begin{center}
{\bf A. Orbital distributions in a core around a SMBH}
\end{center}

In the context of ``isotropic core'' models, we derive the 
distribution of orbital eccentricities that one would measure
at a point inside the core and near to the SMBH.

We assume that $r\ll r_b\ll \rh$, where $r$ is the point
of observation, $r_b$ defines the truncation energy through
$E_b=\phi(r_b)$,  and $\rh$ is the SMBH influence radius.

Assuming a power law in space density outside the core,
the phase space density is given by equation~(\ref{eq:fpl}),
\begin{eqnarray}
f(E) &=& f_0 |E|^{\gamma-3/2}, \ \ E>E_b \nonumber \\
&=& 0, \ \ E\le E_b
\end{eqnarray}
where we have assumed that the gravitational potential
is 
\beq
\phi(r) = -{G\mh\over r}
\eeq
i.e. that $r\ll\rh$.

The velocity space volume element is
\beq
d^3\mathbf{v} = 4\pi v^2 dv \sin\theta d\theta
\eeq
where $\theta$ is the angle between $\mathbf{v}$ and $\mathbf{r}$
and $0\le\theta\le\pi/2$.
Using
\begin{subequations}
\begin{eqnarray}
E &=& {1\over 2} v^2 + \phi(r) = -{G\mh\over 2a}, \\
J &=& rv\sin\theta = \sqrt{1-e^2} {G\mh\over\sqrt{-2E}},
\end{eqnarray}
\end{subequations}
this becomes
\beq
d^3\mathbf{v} = {G^2\mh^2\over 2r^2v_r} {e\over a} da de
\eeq
where $v_r=v\cos\theta$.
The distribution of eccentricities at $r$ is then
given by the integration over $a$, or
\beq
{dN\over de} \propto e \int_{r_b/2}^{r/(1-e)} {da\over av_r} f(a).
\eeq
The lower limit is the semi-major axis of an orbit of energy
$E_b$.
The upper limit on $a$ corresponds to an orbit of eccentricity
$e$ with periapse at $r$.
The integral is zero unless
\beq
{r\over 1-e} > {r_b\over 2}
\eeq
or
\beq
e > 1 - 2{r\over r_b}.
\eeq

Writing
\beq
v_r^2 = {G\mh\over r}\left[2 - {r\over a} - (1-e^2){a\over r}\right]
\eeq
and $f(a) \propto a^{3/2-\gamma}$, this becomes
\beq
{dN\over de} = N_0 e \int_{r_b/2}^{r/(1-e)} da {a^{1/2-\gamma}\over
\sqrt{2 - {r\over a} - (1-e^2) {a\over r}}}, \ \ \ \ e > 1-2{r\over r_b}.
\eeq
Defining the new variable $x = (1-e){a\over r}$ this takes on the
simpler form:
\begin{eqnarray}
{dN\over de} &=& N_0 e\left(1-e\right)^{3/2-0\gamma}\left(1+e\right)^{-1/2}
 \int_{x_1}^1 dx {x^{1-\gamma}\over
\sqrt{\left(1-x\right)\left(x-{1-e\over 1+e}\right)}}, \nonumber \\
x_1 &=& (1-e) {r_b\over 2r}, \nonumber \\
1 &\ge& e \ge 1-{2r\over r_b}.
\end{eqnarray}

\begin{figure}
\includegraphics[clip,width=0.425\textwidth]{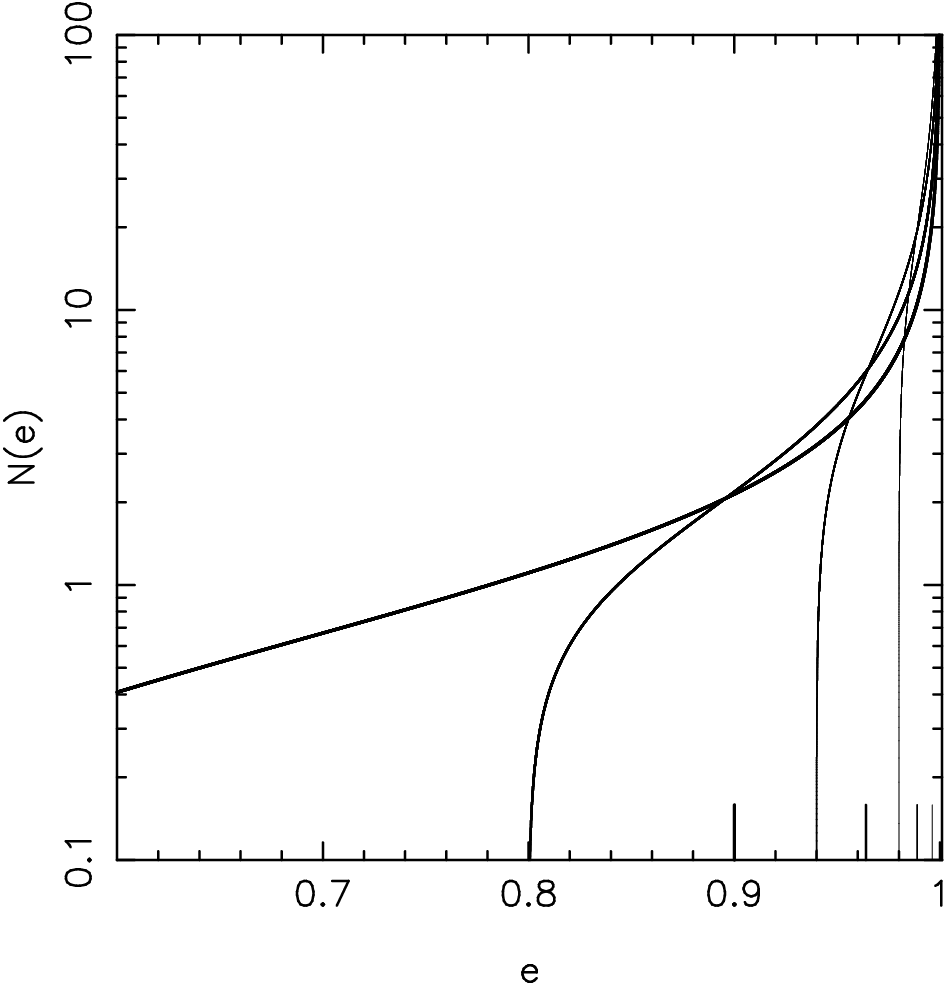}
\caption{Distribution of orbital eccentricities that would
be measured at a given distance from the SMBH in an ``isotropic
core'' model.
The density profile outside the core is $n\propto r^{-1.8}$.
Curves show $N(e)$ at $r=(0.3,0.1,0.03,0.01)\times r_b$;
line width decreases with decreasing $r$.
The vertical tick marks show $\langle e\rangle$ for each curve.
}
\label{fig:dnde}
\end{figure}

Figure~\ref{fig:dnde} shows $N(e)$, normalized to unit total number,
at radii of
\beq
0.3, 0.1, 0.03, 0.01
\eeq
times $r_b$ assuming $\gamma=1.8$.
Also shown are the mean eccentricities at each radius.
$N(e)$ approximates a delta function at $e=1$ for $r\ll r_b$,
since the only orbits that reach into these small radii must
be very eccentric.

Proceeding as before, the distribution of orbital semi-major
axes at $r$ is
\beq
N(a) = 2^{3/2-\gamma} (\gamma-3/2) r_b^{-1}
\left({a\over r_b}\right)^{1/2-\gamma}, \ \ \ \ a \ge r_b/2.
\eeq
where $e=1$ and $r\ll r_b$ have also been assumed.
The mean value of $a$ is 
\beq
{\langle a\rangle\over r_b} = {1\over 2} {\gamma-3/2\over 5/2-\gamma}
\eeq
and for $\gamma=1.8$, 
\beq
{\langle a\rangle\over r_b} = 0.214.
\eeq
Setting $r_b\approx 0.5$ pc, the expected orbit should have
a semi-major axis of $\sim 0.1$ pc and an eccentricity near one.

\end{document}